\documentclass[useAMS,usenatbib,usegraphicx]{aa}

   \usepackage{amsmath}
   \usepackage{amsfonts}   
   \usepackage{amssymb}    
   \usepackage{graphicx}
   \usepackage{subfig}
   \usepackage{natbib}

\include{journaldefs}
\begin{document}

\title{Solar Particle Acceleration at Reconnecting 3D Null Points}
\author{A. ~Stanier
        \inst{1}
	\and
        P. ~Browning
	\inst{1}
	\and
	S. ~Dalla
	\inst{2}}
\institute{Jodrell Bank Centre for Astrophysics, School of Physics and Astronomy, University of Manchester, Manchester M13 9PL, UK \\
 \email{stanier@jb.man.ac.uk}
  \and
Jeremiah Horrocks Institute, University of Central Lancashire, Preston PR1 2HE, UK}

\abstract{The strong electric fields associated with magnetic reconnection in solar flares are a plausible mechanism to accelerate populations of high energy, non-thermal particles. One such reconnection scenario, in a fully 3D geometry, occurs at a magnetic null point. Here, global plasma motion can give rise to strong currents in the spine axis or fan plane. }
{To understand the mechanism of charged particle energy gain in both the external drift region and the diffusion region associated with 3D magnetic reconnection. In doing so we aim to evaluate the efficiency of resistive spine and fan models for particle acceleration, and find possible observables for each.}
{We use a full orbit test particle approach to study proton trajectories within electromagnetic fields that are exact solutions to the steady and incompressible magnetohydrodynamic equations. We study the acceleration physics of single particle trajectories and find energy spectra from many particle simulations. The scaling properties of the accelerated particles with respect to field and plasma parameters is investigated.}
{For fan reconnection, strong non-uniform electric drift streamlines can accelerate the bulk of the test particles. The highest energy gain is for particles that enter the current sheet, where an increasing ``guide field'' stabilises particles against ejection. The energy is only limited by the total electric potential energy difference across the fan current sheet. The spine model has both slow external electric drift speed and weak energy gain for particles reaching the current sheet.}
{The electromagnetic fields of fan reconnection can accelerate protons to the high energies observed in solar flares, gaining up to $0.1$ GeV for anomalous values of resistivity. However, the spine model, which gave a harder energy spectrum in the ideal case, is not an efficient accelerator after pressure constraints in the resistive model are included.}

\maketitle

\section{Introduction}

Observations of Hard X-ray (HXR) and $\gamma$-ray emission from solar flares by the RHESSI space telescope~\citep{lin02} suggest that a large proportion of magnetic energy is converted into kinetic energy of non-thermal accelerated particles. The dominant HXR sources are chromospheric foot-points of the flaring loops, at which there is continuum free-free emission from a beam of energetic electrons in collision with ambient plasma~\citep{brown71}. This continuum spectrum gives beam electron energies from around $10$ keV up to almost $100$ MeV \citep{lin06}. There is line emission at the $\gamma$-ray end of the spectrum from processes involving accelerated ions such as neutron-capture and nuclear de-excitation \citep[see eg.][for a review]{vilmerreview11}, suggesting ions with energies up to $\sim 100$ MeV$/$nucleon. 

When the emission from the foot-points is weak, or when they are occulted by the solar limb, a weaker HXR emission source is sometimes observed above the top of the flare loops \citep{masuda}. Recent observations of two such flares indicate that this emission is non-thermal and that the source is actually the acceleration site for a significant number of energetic electrons~\citep{krucker10,ishikawa11}. The estimated number of energetic electrons is a significant fraction of the emission site density, setting tough efficiency constraints on any proposed acceleration mechanism. 

It is well accepted that magnetic reconnection plays the key role in the dissipation of magnetic energy during a flare and there is a growing body of observational signatures for the process \citep[see][]{mckenzie11}. The site of reconnection in the standard (CSHKP) flare model~\citep[eg.][]{priestflare} is above the thermal loops, not in disagreement with the site of the non-thermal coronal HXR source. Super-Dreicer~\citep{dreicer} electric fields associated with reconnection are one plausible mechanism for particle acceleration and much theoretical work has been done to investigate the efficiency of this mechanism \citep[for review, see][]{zhark2011rev}.

Early work on charged particle trajectories within a reconnecting current sheet concentrated on single particle motion and energy gain in idealised field configurations. \citet{speiser65} found that charged particles in the simplest current sheet, of oppositely directed magnetic field and constant electric field, are trapped so the energy gain is limited only by the sheet length. With an additional small and constant magnetic field component normal to the current sheet plane, the particles are turned by $90^{\circ}$ and ejected from the current sheet into the external drift region. \citet{zhuparks93}, \citet{litvinenkosomov} and \citet{litvinenko96} also included a third component of the magnetic field, a guide field parallel to the electric field. Above a critical guide field the trajectory is stabilised against ejection and the energy gain is once again only bounded by the sheet length~\citep{litvinenko96}.

This early analytical work was extended with 2D (or 2.5D where the fields are invariant in the third dimension) test particle simulations, many of which have been carried out using simple prescribed magnetic and electric fields that would be expected in a reconnection solution to the MHD equations. These simulations consider the effect of the guide field on trajectories and energy spectra within the Harris current sheet \citep{zharkova04,zharkova05,wood05} and magnetic X-points \citep{VB97,hannah,hamilton05}. 

\citet{heerikhuisen02} and \citet{craiglitvin02} used magnetic and electric fields from the exact analytical solutions of \citet{craighenton95} to the 2D incompressible, resistive MHD equations. Also, an approach combining numerical MHD simulations with a test particle code has also been used to study 2D forced reconnection~\citep{gordovskyy2010,gordovskyy10b}. These simulations can include compressibility and time evolution, making them more realistic for coronal plasma. However, analytical solutions are essential to study acceleration due to reconnection in very large Lundquist number plasma at present.

The complexity of the coronal magnetic field in a flaring Active Region motivates the study of test particle motion in fully 3D reconnection geometries. Reconnection models in 3D are comparatively new, but it is clear that there are significant qualitative differences from the familiar 2D models \citep[see eg.][]{pontin10rev}. Reconnection in 3D can occur both with and without magnetic null points. However, the simplest 3D magnetic configuration is based on the potential magnetic field about such a null point. Here, the solenoidal condition defines the magnetic topology of a 1D \textit{spine line} and a 2D \textit{fan surface} \citep[called $\gamma$-line and $\Sigma$-surface by][]{LF90} that separates different magnetic flux domains~\citep[a linear description of magnetic configurations at nulls is given by][]{parnell96}. Although these null points cannot be measured in the corona at present there is some indirect evidence for their existence. Nulls are common features of magnetic topology models that reconstruct the magnetic field from photospheric magnetograms into the corona~\citep[see][for a review]{longcope}. The application of this method at several flare sites suggests the importance of these nulls in certain flares~\citep{desjardins09, aulanier2000, fletcher01}. Reconnection at magnetic null points is also thought to be important for the emergence of new flux from beneath the photosphere into the corona~\citep{torok09,maclean09,liu11}.

The type of reconnection that occurs at a 3D null depends upon the magnetic configuration and global plasma flow. \citet{PT96} proposed two models of reconnection using a potential magnetic field and prescribed global flows that satisfy the ideal MHD equations. In \textit{ideal spine reconnection} a shear flow across the fan plane causes frozen-in flux inflow that converges on the spine axis. At the spine the field reconnects in the presence of singular electric field. For \textit{ideal fan reconnection} the singular electric field occurs in the fan plane, driven by a shearing flow across the spine axis. \citet{craig95}, \citet{CF96} and \citet{CFW97} found exact solutions to the steady and incompressible resistive MHD equations at 3D null points by considering a flux-pileup disturbance field superposed with the background potential magnetic field. The disturbance field induces a current sheet in the spine axis and in the fan plane for the \textit{resistive spine} and \textit{resistive fan} models respectively. \citet{CF98} found corresponding time-dependent solutions to these steady models, and the numerical simulations of~\citet{heerikhuisen04} found reconnection scalings in agreement with both steady state and time dependent models at peak reconnection rate. The 3D MHD simulations of~\citet{pontin07a,pontin07compress} also found a hybrid of the spine and fan models, named \textit{spine-fan reconnection}, when compressibility is included. Recent numerical and analytical study gives additional models for null reconnection when the global plasma motion is rotational rather than a shear flow~\citep[see for review][]{priestpontin09,pontintors11}.

It is not yet known if these null points are effective particle accelerators. Previous work by \citet{db1,db2,db3} and \citet{db4} used the ideal electromagnetic fields of \citet{PT96} in a test particle code, finding the ideal spine reconnection model was effective to accelerate protons and electrons to high energies (max $\sim 10^7$ eV) for solar coronal parameters. The ideal fan reconnection model was less effective for protons, partly as the geometry of the electric drift streamlines was less efficient at delivering particles to regions of high electric field. \citet{guo} used null point magnetic and electric field configurations from MHD simulations, finding that strong electric fields due to convective plasma motion can be efficient at accelerating protons but less so for electrons. \citet{litvinenko06} used the WKB method of \citet{BC88} to show that single protons and electrons close to the null in the reconnecting fan current sheet can achieve the high energies observed in flares. However, this energy is limited as the particles become unstable in the sheet, due to the potential background field, and are ejected.  

In this paper we examine test particle trajectories and energy spectra of protons in electromagnetic fields which are exact solutions to the 3D, steady-state, incompressible and resistive MHD equations at magnetic null points. These are the resistive spine and resistive fan solutions given in~\citet{craig95, CF96} and \citet{CFW97}. We consider trajectories that start both in the outer ideal region, for comparison with particle acceleration results in the ideal models~\citep{db4}, and those that start directly inside the resistive fan current sheet, to compare with the analytical work of \citet{litvinenko06}. 

The paper is organised as follows. In Section~\ref{secmodels} we describe the model fields used, along with parameters chosen considering the pressure constraints and optimisations of~\citet{CFW97} and \citet{cw2000}. We also derive the electric fields and potentials used in the code, and give approximate external drift velocity scalings. In Section~\ref{secresults} we give the results of test particle simulations for thermal distributions of protons starting in the drift region for each model. We choose typical high energy particles from these simulations and follow the single trajectories to understand the energy gain mechanism. We see how drift times and energy spectra scale with the different parameter choices in the fan model. In Section~\ref{secdiscussion} we give a summary and conclusions on the efficiency of each model for accelerating protons.     

\section{Model Fields and the Test Particle Code}\label{secmodels}

We consider two of the reconnection solutions at 3D null points found by~\citet{craig95} and developed in~\citet{CF96} and~\citet{CFW97}. The fields satisfy the resistive, steady-state, and incompressible MHD equations. These are normalised in the usual way by the characteristic magnetic field strength $B_0$, at a typical length scale $L_0$ from the null point, and by density $\rho_0$. This choice leads to the natural units for velocities in terms of the external Alfv\'en speed, $v_0 = v_{Ae} =B_0/\sqrt{\rho_0\mu_0}$. The thermal pressure is normalised by the dynamic pressure at Alfv\'en speed, $p_0 = \rho_0 v_{Ae}^2$, so that the dimensionless pressure on the $L_0$ boundary is half the plasma beta $p_e = \beta_e/2$. The dimensionless resistivity, $\eta$, is given by
\begin{equation}\label{eta}\eta = \frac{\eta^d}{L_0v_{Ae}\mu_0} \equiv S^{-1},\end{equation}
where $\eta^d$ is the dimensional resistivity (Spitzer resistivity in the case of purely collisional plasma), $\mu_0$ is the magnetic permeability, and $S$ is the Lunquist number which is typically very large in the solar corona $S \sim 10^{12}- 10^{14}$. 

For completeness, the main properties of the solution are given here; see \citet{CF96} and \citet{CFW97} for more detail. After normalisation the governing equations consist of the momentum equation, which in curled form is
\begin{equation}\label{momeqn}(\boldsymbol{u}\cdot \boldsymbol{\nabla}) \boldsymbol{\omega} - (\boldsymbol{\omega} \cdot \boldsymbol{\nabla})\boldsymbol{u} = (\boldsymbol{B} \cdot \boldsymbol{\nabla}) \boldsymbol{J} - (\boldsymbol{J} \cdot \boldsymbol{\nabla}) \boldsymbol{B},\end{equation}
and the induction equation,
\begin{equation}\label{indeqn}(\boldsymbol{u}\cdot\boldsymbol{\nabla})\boldsymbol{B} - (\boldsymbol{B}\cdot \boldsymbol{\nabla})\boldsymbol{u} = \eta \nabla^2 \boldsymbol{B},\end{equation}
with the solenoidal and incompressibility conditions,
\begin{equation}\boldsymbol{\nabla} \cdot \boldsymbol{B} = 0,\quad \boldsymbol{\nabla}\cdot \boldsymbol{u} = 0.\end{equation}

Here $\boldsymbol{J}$ is the current density and $\boldsymbol{\omega}$ is the vorticity in terms of the bulk plasma velocity $\boldsymbol{u}$. In this normalised form they are
\begin{equation}\boldsymbol{J} = \boldsymbol{\nabla} \times \boldsymbol{B}, \quad \boldsymbol{\omega} = \boldsymbol{\nabla} \times \boldsymbol{u}.\end{equation}

The three dimensional analytic solutions of~\citet{craig95},~\citet{CF96} and~\citet{CFW97} have magnetic and flow fields of the form
\begin{equation}\boldsymbol{B} = \lambda \boldsymbol{P} + \boldsymbol{Q},\end{equation}
\begin{equation}\boldsymbol{u} = \boldsymbol{P} + \lambda \boldsymbol{Q},\end{equation}
where the scalar $0\le\lambda < 1$ gives the shear between the $\boldsymbol{B}$ and $\boldsymbol{u}$ fields. The vector field $\boldsymbol{P}(x,y,z)$ is a potential background field of strength $\alpha$, and $\boldsymbol{Q}$ is a disturbance field of strength $B_s$ which gives rise to current in the models.

For comparison with the particle acceleration results at 3D nulls in ideal MHD \citep{db1} we choose the z-axis to be aligned with the spine, with $z=0$ as the fan plane. It must be noted that this choice of axis differs from that used by \cite{CFW97}. We study only the \textit{proper radial null}~\citep{PT96} where the background magnetic field lines in the fan plane lie in the radial direction. This background field is then written as
\begin{equation}\label{potfield}\boldsymbol{P} = \frac{\alpha}{2}(x\boldsymbol{\hat{x}} + y\boldsymbol{\hat{y}} -2z\boldsymbol{\hat{z}}),\end{equation}
with $\alpha$ giving the sign and strength of the field. For the spine model, the displacement field distorts the fan plane in the z-direction $\boldsymbol{Q}_S = Z(x,y)\boldsymbol{\hat{z}}$. For the fan model, it distorts the spine in the x-direction $\boldsymbol{Q}_F = X(z)\boldsymbol{\hat{x}}$ (the more general fan case given in \cite{CFW97} of $\boldsymbol{Q}_F = X(z) \boldsymbol{\hat{x}} + Y(z) \boldsymbol{\hat{y}}$ has not been covered here).

\subsection{Spine Analytic Fields}

The disturbance field for the spine model in cylindrical co-ordinates $(r,\phi,z)$ is
\begin{equation}\label{spinedisp}\boldsymbol{Q}_S = Z(r,\phi)\boldsymbol{\hat{z}} = \frac{B_s r}{r_\eta}\, M\left(\frac{3}{2},2,-\frac{r^2}{r_\eta^2}\right) \sin(\phi)\boldsymbol{\hat{z}},\end{equation}
\citep{CFW97} in terms of the confluent hypergeometric (Kummer) function $M(a,b,\zeta)$ \citep[][]{abramowitz}. The flux pile-up factor $B_s$ gives the approximate strength of the magnetic field at a dimensionless distance $r_\eta$ from the spine axis, where $r_\eta$ is defined as 
\begin{equation}r_\eta \equiv \sqrt{4\bar{\eta}} \equiv \sqrt{\frac{4\eta}{|\alpha| (1-\lambda^2)}}.\end{equation}
It is the radius of a cylindrical region centred on the spine axis where resistive effects become significant~\citep{CFW97}. 

\begin{figure}
\centering
\subfloat[]{\label{spineBline:a}
\includegraphics[width=0.5\textwidth]{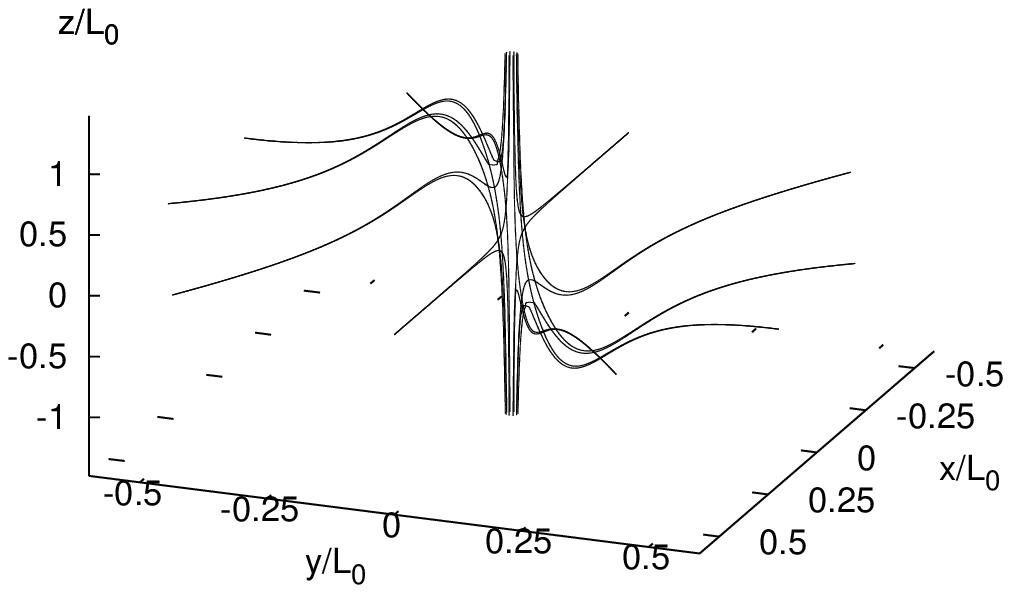}}\\[-0.2cm]
\subfloat[]{\label{spineBline:b}
\includegraphics[width=0.5\textwidth]{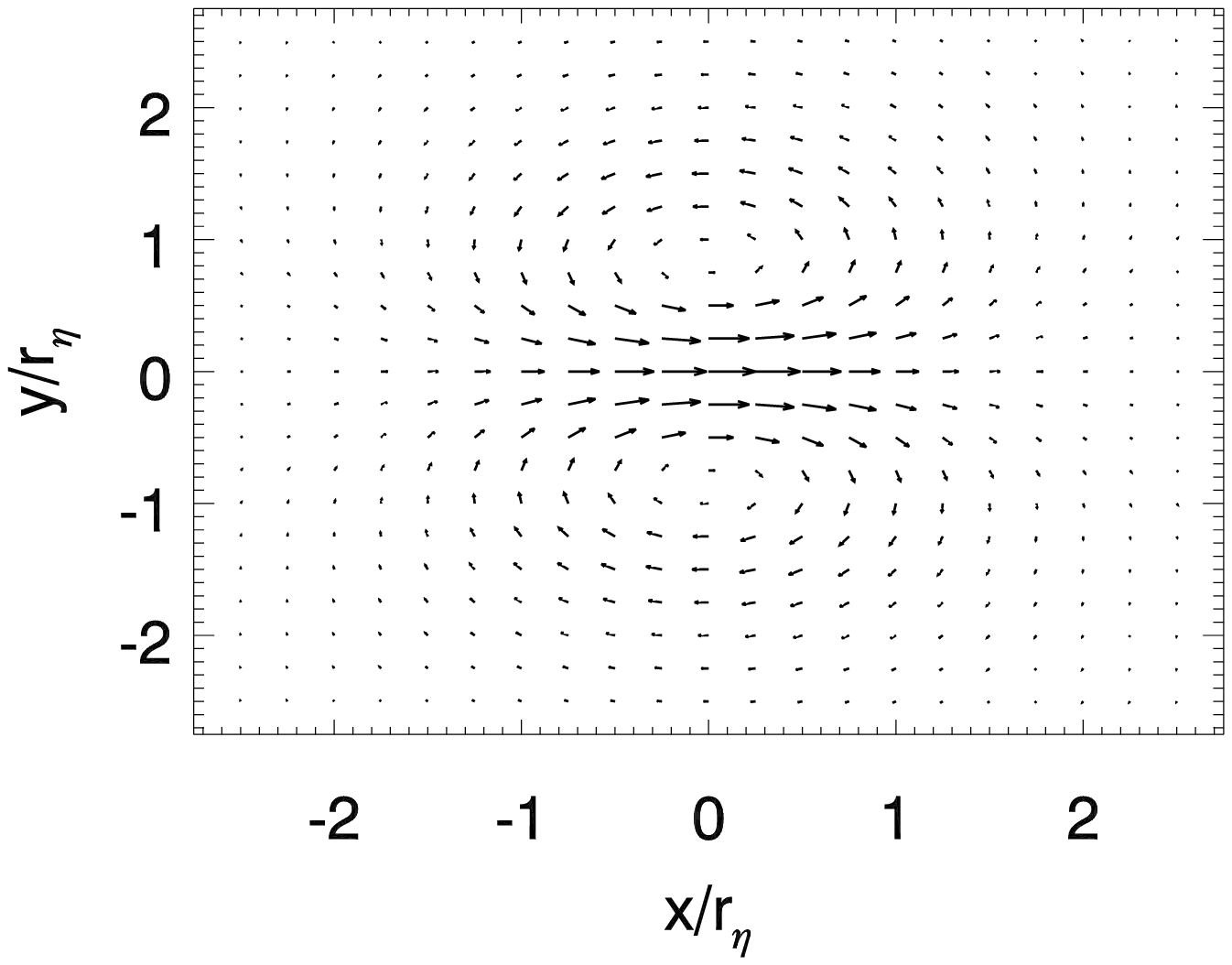}}
\caption{a) Representative magnetic field lines for the spine model with parameters $\lambda = 0.75$, $B_s = 3.4$, $\alpha = -2$, $\eta = 3 \times 10^{-3}$. The field lines are seeded from the top and base of the spine axis. b) Showing the direction and relative strength of the current in a plane of constant z for the same parameters. Here, $r_\eta=\sqrt{4\bar{\eta}}\approx 0.12$ is the size of the resistive region centred on the spine axis.}
\label{spineBline}
\end{figure}

The form of the displacement field in equation~(\ref{spinedisp}) is only a solution to the governing equations provided $\alpha < 0$. This gives frozen-in plasma inflow along the fan plane converging on the spine and outflow in the $\pm z$ directions away from the null point. The magnetic field in the outer (ideal) region is also directed inwards along the fan plane and outwards along the spine axis. Some representative magnetic field lines are shown in Figure~\ref{spineBline}\subref{spineBline:a}, the displacement term shears the fan plane at $\phi=\pm \pi/2$ towards the spine axis while the fieldlines in $\phi = 0,\pi$ of the fan plane remain perpendicular to the spine.  

To integrate particle trajectories using a test particle model we require the electric field. We calculate this from the uncurled form of equation~(\ref{indeqn}) as
\begin{equation}\boldsymbol{E}(r,\phi) = \frac{\eta}{r}\frac{\partial Z}{\partial \phi} \boldsymbol{\hat{r}} + \left[(1-\lambda^2)P_r Z - \eta \frac{\partial Z}{\partial r}\right]\boldsymbol{\hat{\phi}},\end{equation}
where $P_r = \alpha r/2$ is the radial part of the potential field.

This electric field is curl-free (as required for steady-state) and so we can calculate the electric potential $V$ to use as a check of energy conservation. This can be found by integrating $\boldsymbol{E} = - \boldsymbol{\nabla}V$ to get
\begin{equation}\label{potspine}V(r,\phi) = \cos{\phi}\left[\frac{\alpha\, (1-\lambda^2)\,r^2\,f(r)}{2} - \eta\,r\,f'(r)\right],\end{equation}
where $f(r)$ is the radial part of the displacement field in~(\ref{spinedisp}), $Z(r,\phi) = f(r)\sin{\phi}$.

We can study the behaviour of these fields at small and large distances using the truncated power series and asymptotic formulae for the Kummer function respectively \citep{abramowitz}. For all our cases the third argument in the Kummer function is negative. We have for $0 \le \xi \ll 1$,
\begin{equation}\label{powerseries}M(a,b,-\xi) \approx 1 - a\xi/b,\end{equation}
and for $\xi \gg 1$,
\begin{equation}\label{asympspine}M(a,b,-\xi) \approx \frac{\Gamma(b)}{\Gamma(b-a)}\xi^{-a},\end{equation}
in terms of the Gamma function $\Gamma(b)$. Near the spine axis $r\approx 0$, the only contribution to the electric field is from current in the x-direction and
\begin{equation}\boldsymbol{E}(r\ll r_\eta)\approx\eta \boldsymbol{J}(0) = \frac{\eta B_s}{r_\eta} \boldsymbol{\hat{x}}.\end{equation}
The full current distribution is plotted in Figure~\ref{spineBline}\subref{spineBline:b}, it forms two cylindrical vortex structures that are localised with respect to the resistive region and invariant in the z-direction.
 
At large distances from the spine, the electric field goes as
\begin{equation}\label{spineasymptE}\boldsymbol{E}(r\gg r_\eta) \approx \frac{-2\eta B_s}{\sqrt{\pi}}\frac{\sin{\phi}}{r}\boldsymbol{\hat{\phi}}.\end{equation}
This has the same functional form as the ideal spine solution of \citet{PT96}, the subject of previous work on 3D null-point particle acceleration by \citet{db1,db2} and \citet{db4}. Indeed, by simple choice of parameters we could set the magnetic and electric fields to asymptotically match those of the ideal case. Some care must be taken here as \cite{db1} studied \textit{positive nulls}, where $\alpha>0$, and with $E(0<\phi<\pi)>0$. We have opposite sign for both electric and magnetic fields giving the same electric drift inflow quadrants but different sign for the convective electric field. Particles that become non-adiabatic in the external region $r \gg r_\eta$, and gain energy parallel the electric field, will rotate about the spine in the opposite direction to those in \citet{db1,db2}. In this paper we will only qualitatively compare particle trajectories in the ideal and resistive spine models as an asymptotic match will give rise to unphysical hydromagnetic pressures on the edge of the resistive region $r\approx r_\eta$ that were absent in the simplified ideal model (see below). 

The thermal pressure profile for the spine model can be found from integrating the uncurled form of equation~(\ref{momeqn}). It is given in \cite{CFW97} to be 
\begin{equation}\label{spinepressure}p = p_0 - \frac{1}{2}\left(P^2 + Z^2\right) + \lambda\alpha z Z,\end{equation}
where $p_0$ is the gas pressure at the null point, the first term inside the brackets is due to dynamic pressure from the background flow and the other two terms are from balance with magnetic pressure. All terms except for $p_0$ are negative, as $\alpha<0$, so constraints must be put on the values of $\alpha$ and $B_s$ in order to avoid unphysical negative pressures as discussed in \citet{lfp96,litvincraig99, CFW97} and \citet{cw2000}. We give some of the arguments here for the sake of completeness (see above references for more detail).

 The strong electric field (fast electric drift) simulations for the ideal spine model studied by~\citet{db1} were characterised by the dimensional value of the electric field $E_0=1500$ $V/m$ on the $r=1$, $\phi=\pi/2$ boundary (or normalising by suitable solar coronal values, $v_{Ae} = 6.5\times 10^6$ ms$^{-1}$, $B_0 = 0.01$ T, gives $E \approx 1/40$). This value can be equated with equation~(\ref{spineasymptE}). Crucially, to match the external electric field in the resistive model to the fixed amplitude electric field in the ideal spine reconnection model requires the scaling $B_s \sim \eta^{-1}$ as $\eta$ is reduced to suitable solar coronal values (\citet[][]{CFW97} showed that if we require displacement field at the boundary $Z(1,\pi/2) \sim 1$ this also gives $B_s \sim \eta^{-1}$). However, this scaling gives rise to large magnetic pressure on the sheet edge. The maximum of the displacement field occurs at $r\approx r_\eta$ where $Z(r_\eta) \sim B_s$ giving magnetic pressure there from equation~(\ref{spinepressure}) $Z^2 \sim B_s^2 \sim \eta^{-2}$. To avoid negative thermal pressure in the model this requires the null point pressure $p_0 > (Z(r_\eta))^2 \sim \eta^{-2}$ which is unphysically large for the values of $\eta$ considered.

 \citet{CFW97} showed that $B_s$ must be limited to a saturation value on $r=r_\eta$, giving weak electric fields and small amplitude displacement field on the boundary $Z(1) \ll 1$. Also, at $r=1$, $z \ll 1$ we have dynamic pressure due to bulk fluid inflow $p \approx p_0 - P^2$ where $P(1) \sim \alpha$. We must constrain $\alpha \le B_s$ or this dynamic pressure will require the gas pressure at the null to be even larger. The maximum value we can take for $p_0$ is the largest possible external hydromagnetic pressure $p_{e,max}$ available to drive the reconnection. We follow \cite{CFW97} and take $p_{e,max} = B_{e,max}^2/2$ where the maximum external magnetic field is that of a sunspot at the photosphere, $B_{e,max}=0.3$ T. This gives a normalised saturation value $B_{s,max} = 30$.

So far we do not know the value $\alpha$ should take, but expect that the bulk fluid exhaust from the reconnection region is of the order of the local Alfv\'en speed. The exhaust on the edge of the current sheet at a global distance from the null, $r=r_\eta$, $\phi=\pi/2$, $z=1$, is given by 
\begin{equation*}|v(r_\eta,\tfrac{\pi}{2},1)| \approx \lambda B_s - \alpha,\end{equation*}
where the local Alfv\'en speed is 
\begin{equation*}|v_A(r_\eta,\tfrac{\pi}{2},1)| = |B(r_\eta,\tfrac{\pi}{2},1)|  \approx B_s -\lambda \alpha\end{equation*}
for our choice of normalisation. As we are not interested in the case where $\lambda=1$ (where there is no shear between the velocity and magnetic fields) we have $\alpha \approx - B_s$ for Alfv\'enic exhaust. This is the largest magnitude of $\alpha$ we can take without having problems due to dynamic pressure. It also leads to the thinnest current sheet and thus maximises the current density in the resistive region. However, as \citet{cw2000} show, the dissipation rate is
\begin{equation}W_\eta = \eta \int J^2 dV \approx \pi \eta B_s,\end{equation}
which has no $\alpha$ dependence as the increase in current density due to resistive region thinning is cancelled by the $r_\eta^2$ dependence of the total dissipation volume.

The electric drift velocity in the external region is given by
\begin{equation}\label{spinedriftext}v_E(r\gg r_\eta) \approx \frac{\eta B_s \sin{\phi}}{\lambda |\alpha| \sqrt{\pi}}\left(\frac{-2z\boldsymbol{\hat{r}} -r \boldsymbol{\hat{z}}}{r(r^2/4 + z^2)}\right)\quad [v_{Ae}]\end{equation}
which is very slow when $|\alpha| = B_s$. It is thus necessary to limit the magnitude of $\alpha$ so that results can be obtained with reasonable integration times. For the simulations in Section~\ref{secresults} we use $B_s=10$, $\alpha=-0.1$, this limits the reconnection exhaust close to the spine current sheet to sub-Alfv\'enic speeds.

\subsection{Fan Analytic Fields}

\begin{figure}
\centering
\includegraphics[width=0.5\textwidth]{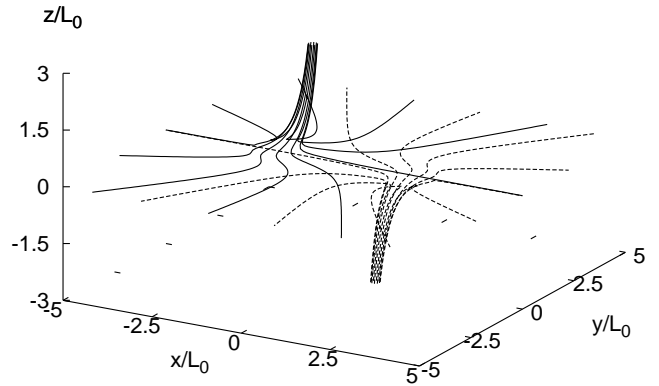}
\caption{Representative magnetic field lines for the fan solution with parameters $\lambda = 0.75$, $B_s = \alpha = 10$, $\eta = 10^{-6}$. The lines are again seeded from the top (solid lines) and base (dashed lines) of the spine.}
\label{fanBline}
\end{figure}

The displacement field for the fan model is
\begin{equation}\label{fandisplacement} \boldsymbol{Q}_F = X(z)\boldsymbol{\hat{x}} = \frac{B_{s}\,z}{\bar{\eta}^{1/2}}M\left(\frac{3}{4},\frac{3}{2},\frac{-z^2}{2\bar{\eta}}\right)\boldsymbol{\hat{x}},\end{equation}
\citep{CFW97}.
We define $z_\eta$ as
\begin{equation}z_\eta \equiv \sqrt{2\bar{\eta}} \equiv \sqrt{\frac{2\eta}{\alpha(1-\lambda^2)}},\end{equation}
the approximate height at which $X$ takes the maximum value, $X_{max} \approx B_s$. It is a measure of the height of a resistive region centred on the fan plane, $z=0$. This form of solution is only valid for $\alpha>0$ which gives a positive null point, the field is washed in from the global boundaries at $z=\pm 1$ and it exits the simulation box radially along the fan plane. Some representative magnetic field lines are shown in Figure~\ref{fanBline}; the displacement field shears the spine axis as it approaches the fan plane giving rise to strong current inside the resistive region.

The electric field is
\begin{equation}\label{fanelec}\boldsymbol{E} = \boldsymbol{\hat{y}}\left[\eta \, X'(z) - (1-\lambda^2)P_z X(z)\right] + \boldsymbol{\hat{z}}\left[(1-\lambda^2)P_yX(z)\right],\end{equation}
and the electric potential is
\begin{equation}\label{potfan}V(y,z) = -\alpha y (1-\lambda^2) \left[\bar{\eta}X'(z) + zX(z)\right],\end{equation}
where $\boldsymbol{J}(z=0) = X'(0)\boldsymbol{\hat{y}}$ is current density at the centre of the sheet. The current only has z-dependence; it is infinite in extent in the x and y directions. This is clearly unrealistic, although resistive MHD simulations by \citet{pontin07compress} find that \textit{spine-fan} reconnecting current sheets formed due to shear flows around a null point spread out along the fan plane in the incompressible limit. Note that analytic multiple null solutions found by \citet{craig99multiple} have finite current sheets, avoiding this problem. In our simulations below we consider particle acceleration only within a restricted range of $5\, L_0$, effectively limiting the size of the current sheet.

The thermal pressure profile for the fan model is~\citep{CFW97},
\begin{equation}p = p_0 - (P^2 + X^2)/2 - \alpha \lambda x X/2.\end{equation}
However, in this case a displacement field of order unity on the $z=1$ boundary, $X(1) \sim 1$, gives the scaling $B_s \sim \eta^{-1/4}$~\citep{CFW97}. This gives much weaker hydromagnetic pressure on the current sheet edge compared to the spine model but it is still too large for the values of $\eta$ considered. Again we saturate $B_{s,max} = 30$ and we have $\alpha \le B_s$ to avoid problems from dynamic pressure. 

\citet{cw2000} show that the Ohmic dissipation rate from the fan model is
\begin{equation}W_\eta = \eta \int J^2 dV \sim \eta B_s^2/z_\eta\end{equation}
and so in this case, for fixed (saturated) $B_s$, the maximum dissipation occurs with the thinnest current sheet (the so called optimised solution). The thinnest sheet we can have subject to the dynamic pressure constraint is when $\alpha = B_s$ (for any fixed value of $\lambda$). Also, as this choice gives the largest current density, it maximises the resistive electric field within the sheet which is interesting for particle acceleration. As above, this choice of $\alpha$ sets the bulk fluid exhaust at $x^2 + y^2 = 1$, $z=z_\eta$ to the local Alfv\'en speed. 

Using the asymptotic approximation~(\ref{asympspine}) we find that the z-component of the electric drift that brings the particles to the fan plane is, for $x,\,z \gg \eta^{1/2}$,
\begin{equation}\label{fanexternaldrift}v_{Ez} \approx \frac{(1-\lambda^2)P_xP_zX(z)}{\lambda P^2} \sim B_s^{3/4}\eta^{1/4},\end{equation}
for the optimised solution $\alpha=B_s$. This gives electric drift inflow for positive $x$, $z$ (as $P_z<0$), and outflow for positive z and negative x. It is much faster than the spine case due to the more favourable scaling with resistivity. There are also fast drift streamlines in the x-y plane that that can be found from the numerical (or approximate analytical) solution of
\begin{equation}\frac{dx}{v_{Ex}} = \frac{dy}{v_{Ey}},\end{equation} 
we numerically plot these streamlines on top of the single particle trajectory results.

\subsection{Test Particle Code and Parameter Choice}

We modify the test particle code of \citet{db1,db2,db3} and~\citet{db4} to use the electromagnetic fields given above~\citep[from the solutions of][]{CFW97}. A Variable-Step Variable-Order Adam's method, where the step size is recalculated to properly resolve gyro-motion, is used to integrate the relativistic Lorentz equation.
\begin{equation}\frac{d\boldsymbol{p}}{dt} = \frac{q}{m} \left(\boldsymbol{E} + \frac{\boldsymbol{p}}{\gamma m} \times \boldsymbol{B}\right),\end{equation}
where $\boldsymbol{p}$ is the momentum of the particle, $\gamma$ is the Lorentz factor, $q$ and $m$ are the charge and rest mass and $\boldsymbol{E}$ and $\boldsymbol{B}$ are the analytic expressions for the electric and magnetic fields for each model.

We use the expressions for the electric potential $V$, calculated in equations~(\ref{potspine}) and~(\ref{potfan}) to calculate the electric potential energy at each time step. With this we verify that the total energy
\begin{equation}W = \epsilon_k + qV\end{equation}
is conserved where $\epsilon_k = (\gamma - 1)m c^2$. For each simulation we find that this is conserved up to 5 significant figures. Also, to check the code handles non-adiabatic motion in strong magnetic field gradients of a current sheet we reproduce the results of~\citet{speiser65}, including the ejection time for the case with background field.

We choose $L_0$, the normalising length scale, to be $L_0=10^4$ m for global simulations to keep integration times short. This size of simulation box can be considered as a local region around the null at which the linear background field and flow in equation~(\ref{potfield}) are good approximations. We use a larger value of $L_0 = 10^6$ m for simulations where particles are initially distributed within the current sheet, as velocities are typically much faster here. Note that a change in $L_0$ also changes the value of $\eta$ as given in equation~(\ref{eta}).

All magnetic fields mentioned in Section~\ref{secmodels} have dimensions of $B_0 = 0.01$ T, typical for the solar corona. We set $v_0 = v_{Ae} = 6.5 \times 10^6$ ms$^{-1}$ (corresponding to a number density $n_0 = 1.126 \times 10^{15}$ m$^{-3}$). All dimensionless times quoted are in terms of the gyro-period, $T_{\omega} = 2\pi m/(q B_0)$.

To examine single particle trajectories in both models we choose values of $B_s=10$, $\lambda=0.75$, $\eta = 10^{-6}$. This $\eta$ value is rather large, towards the highest possible anomalous resistivities (with $L_0 = 10^4$ m), but is useful to observe particles entering the current sheet. We vary all three of these parameters to produce scalings of energy gain and drift times, where we use values as low as those expected in purely collisional plasma i.e. $\eta=10^{-12}$.

\section{Results}\label{secresults}

\subsection{Spine Global Trajectories}

Initially, we place a distribution of $5\,000$ protons with Maxwellian velocities of temperature $T=10^6$ K (86 eV) in the spine model fields. The protons have positions from a uniform random distribution at a global distance $x^2 + y^2 + z^2 = 1$ from the null point. We only discuss here protons that start in the upper right inflow region of longitude $0<\phi<180^{\circ}$ and latitude $0<\beta <90^{\circ}$ (here $\phi=0$ is the x-axis and $\beta = 0$ is the fan plane). 

\begin{figure}
\includegraphics[width=0.5\textwidth]{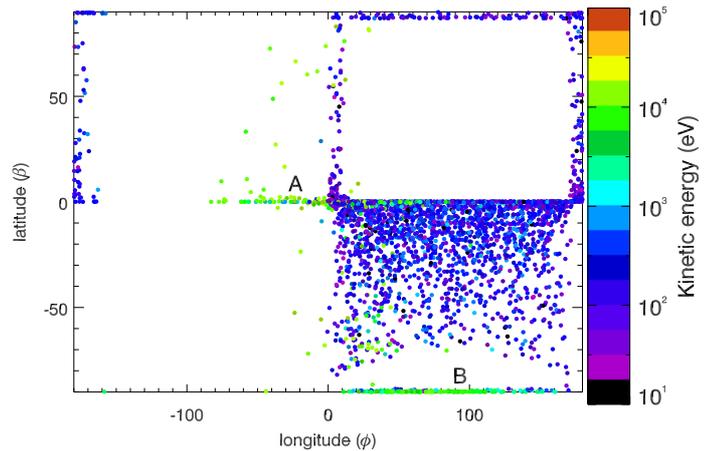}
\caption{Angular distribution of protons from the null point for spine model at $t=1.6 \times 10^6 T_{\omega,p}$, at which the energy spectrum is steady state, for parameters $\lambda=0.75$, $B_s=10$, $\alpha=-0.1$, $\eta=10^{-6}$. The initial distribution was Maxwellian at $T=86$ eV in the upper right inflow region.}\label{spineangular}
\end{figure}

\begin{figure}
\includegraphics[width=0.5\textwidth]{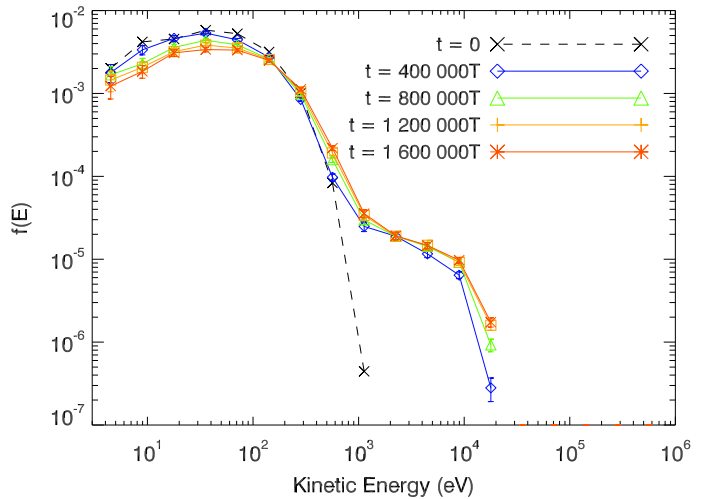}
\caption{Energy spectrum from the many particle simulation for protons in the spine model, with parameters $\lambda=0.75$, $B_s=10$, $\alpha=-0.1$, $\eta=10^{-6}$. For particles leaving the $R=5$ sphere, the energy at the time of crossing is used.}\label{spine_global_energy}
\end{figure}

Figure~\ref{spineangular} shows the final spatial distribution and energies of the particles at time $t=1.6 \times 10^6\,T_{\omega,p} \approx 10$ s, at which the energy spectrum in Figure~\ref{spine_global_energy} becomes steady-state. Those protons starting in the lower left inflow region have final distributions as in Figure~\ref{spineangular} after reflections in both $\phi=0$ and $\beta=0$ apart from statistical differences. The parameters used here are $B_s = 10$, $\eta = 10^{-6}$, $\alpha = -0.1$. This value of $\alpha$ limits the bulk flow exhaust speed to be sub-Alfv\'enic but it increases the electric drift speed in the external region (see equation~(\ref{spinedriftext})) due to weaker magnetic field on $r=1$. This gives reasonable simulation times, but there are still some particles in the upper right inflow quadrant at the end of the simulation.

There are two main populations of accelerated particles. The population labelled 'A' in Figure~\ref{spineangular} is close to the fan plane, $|\beta| \lesssim 10^{\circ}$, with energy $\epsilon_k \gtrsim 1$ keV, and with longitude $-90^{\circ}\lesssim \phi \lesssim 90^{\circ}$ comprising of about $8\%$ of the total proton number. The maximum particle energy of this population is about $15$ keV. Note that the current in the spine axis is aligned with $\phi=0$ through the centre of this population. There are also some high energy protons scattered at large positive latitudes for $\phi \lesssim 0$, and at large negative latitudes for $\phi \gtrsim 0$. To look more closely at what is happening here we will choose a typical proton from this population and follow its trajectory below.

For those particles that have crossed the fan plane, $\beta = 0$, into the lower right outflow quadrant, the spatial and energy distribution looks similar to the ideal spine case in \citet{db2}. The accelerated population which has $\epsilon_k \gtrsim 1$ keV and $\beta \lesssim -85^{\circ}$ is labelled 'B'. This population is about $6\%$ of the total protons in the simulation and the maximum kinetic energy in this population is $\epsilon_{k,max} \approx 12$ keV. The angular distribution differs slightly with the ideal case in that there are few particles found between the latitudes $-70^{\circ}<\beta<-85^{\circ}$; particles appear to be closer to the negative spine axis in the resistive case.

\begin{figure}
\includegraphics[width=0.5\textwidth]{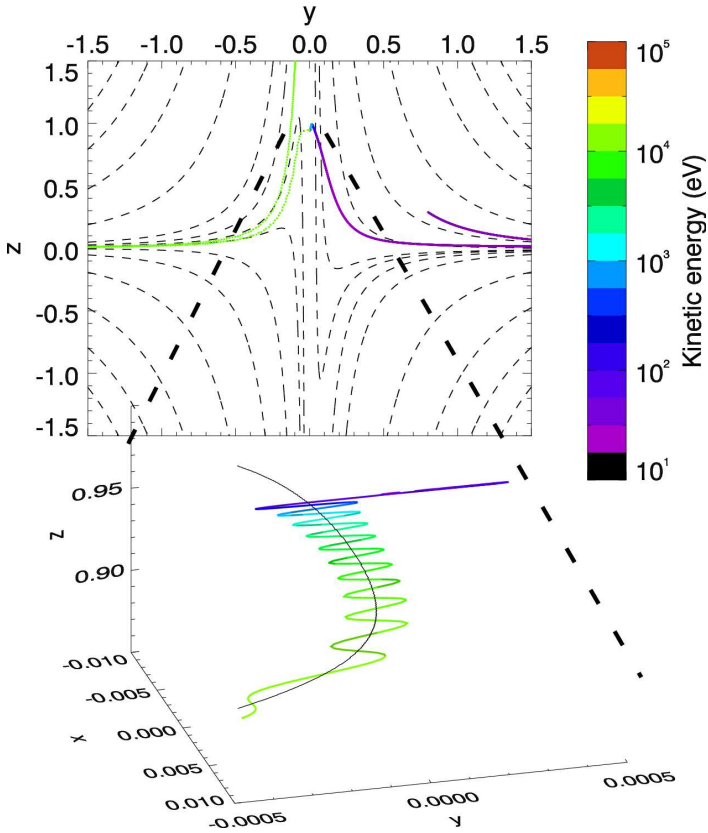}
\caption{Typical proton global trajectory from population 'A' for parameters $\lambda=0.75$, $B_s=10$, $\alpha=-0.1$, $\eta=10^{-6}$. The particle is taken from the many particle simulation having initial position $x=(-0.52,0.80,0.29)$ and velocity $v=(-0.0044,0.0013,-0.0088)v_{Ae}$. The magnetic field lines (thin dashed) are a projection of the field from the plane of the trajectory $\phi \approx 120^{\circ}$. Inset shows the 3D trajectory of the proton as it crosses the spine-axis, the solid line in the centre is the line $B_z(x,y,0.95)=0$.} \label{spinecross}
\end{figure}

A typical proton trajectory from population 'A' is shown in Figure~\ref{spinecross}. The proton which starts at $(x_0,y_0,z_0)=(-0.52,0.80,0.29)$ in the upper right hand inflow quadrant initially moves away from the null but mirror bounces and travels back towards the spine along the fan plane. The electric drift speed increases towards the spine causing the proton to enter the resistive region, which has radius $r_\eta \approx 0.01$, about the spine axis. It enters at $(x,y,z) \approx (-0.01,0,0.95)$ after $t=3 \times 10^5\, T_{\omega,p} \approx 2$ s (inset). At this point the proton becomes unmagnetised as the gyro-radius becomes comparable to the length-scale of magnetic field gradient $\rho/L_{\nabla B}>1$ (we typically find that gyro-motion starts to break down when $\rho/L_{\nabla B} \gtrsim 10^{-2}$).

\begin{figure}
\includegraphics[width=0.5\textwidth]{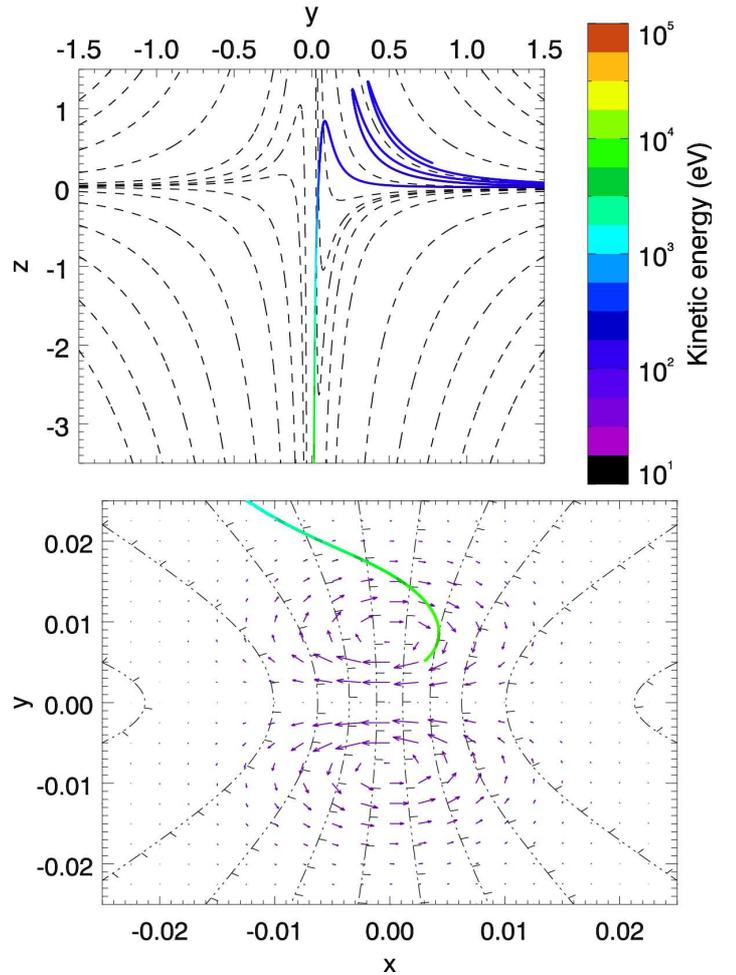}
\caption{Typical proton trajectory from population 'B' in the many particle simulation, with initial position $x=(-0.54, 0.78, 0.31)L_0$ and velocity $v=(-0.004,-0.006,0.002)v_{Ae}$. The dashed lines show the projection of the magnetic field from the plane of motion, $\phi\approx 110^{\circ}$, onto the y-z plane. Inset shows the motion in the x-y plane close to the spine axis. The purple arrows show the direction and relative magnitude of the gradient drift velocity and the dash-dotted lines show contours of the electric potential, with the intersecting tick mark indicating lower potential to the right.}\label{basespinetraj}
\end{figure}

The proton is then directly accelerated in the x-direction parallel to the current at the spine with $du/dt \approx q E_0/m$ as it crosses $r=0$. For small displacements in the y-direction a strong Lorentz force due to the $B_z$ field returns it to $y=0$ line. These oscillations are Speiser-like \citep{speiser65} with frequency approximately $\omega \propto t^{1/2}$.

The Speiser-like motion finishes and the first gyrations start (not shown) when the proton reaches $r \approx 5r_\eta$ at which $\rho/L_{\nabla B} \lesssim 1$. However, the energy gain of $\epsilon_k \approx 11$ keV is localised to within $x \approx 2r_\eta$, during which the trajectory does not deviate much from the x-direction (note the y-axis scale in the inset of Figure~\ref{spinecross}). In effect, the proton has left the localised current sheet while unmagnetised but before it can be ejected by the background field components, in contrast to 2D current sheet configurations with weak guide field \citep[eg.][]{speiser65,litvinenko96}. Figure~\ref{spinecross} may give the impression that the particle is being ejected, however, this is just the centre of the Speiser-like oscillations following the $B_z(x,y,z=const.)=0$ line (which here is not straight as in the usual 2D configurations). This behaviour is evident considering the $\boldsymbol{F} =  q\boldsymbol{v} \times \boldsymbol{B}$ force for the unmagnetised proton if $B_z$ is the dominant component of the magnetic field.

After the proton becomes re-magnetised at $r\approx 5r_\eta$ it has weak electric drift, $v_{E} \ll v_{\omega}$. It follows the fieldlines closely and mirror bounces travelling back towards the spine: there the proton is taken up to high latitude before it bounces again. This mirror bouncing is the reason for the 'scattered' accelerated protons in Figure~\ref{spineangular}, some of which are at large latitudes.

A typical particle trajectory chosen from population 'B' is shown in Figure~\ref{basespinetraj}. The proton starts at $(-0.54, 0.78, 0.31)$ and drifts towards the spine but bounces and crosses the fan plane instead. It exits the simulation box down the base of the spine axis, reaching an energy $\epsilon_k = 6.72$ keV as it crosses $z=-5$. As there is no electric field in the z-direction, the energy gain must occur due to motion in the x-y plane, which is also shown in Figure~\ref{basespinetraj}. The proton enters the region close to the spine axis parallel to a contour of the electric potential, but then drifts across the contour due to strong gradient drift. While the proton gains energy, the gradient drift is larger than the electric drift by a factor of 2 with the latter directed inwards towards the current sheet. The proton is stopped as it reaches $z= -5L_0$ which we do consistently throughout these simulations. At the time of stopping it is actually losing energy as it re-crosses the same electric potential contours. However, some other protons from the many-particle simulation in Figure~\ref{spineangular} reach the current sheet at low latitudes, gaining higher energy.

The energy spectrum for the spine simulation is shown in Figure~\ref{spine_global_energy}. If protons cross the $R=5 L_0$ spherical boundary we use the energy at the instant of crossing (if this is not done some protons reach order $\sim 10^2 L_0$ which becomes unrealistic as the background field increases without bound away from the null, also causing the time-step to decrease and simulation time to increase). The initial Maxwellian spectrum hardens to a broken power law with maximum energy of about $\epsilon_k \approx 15$ keV. This maximum energy can be understood as the difference in potential energy across the spine current sheet, $\epsilon_k \sim q E x_{acc}$ where $E \approx E_0 \approx \eta B_s/r_\eta \,\, [v_{Ae} B_0]$ and $x_{acc} \approx 2 r_\eta \,\, [L_0]$ is the acceleration distance (from $-r_\eta \lesssim x \lesssim r_\eta$), as the electric field drops off quickly for $|x|>r_\eta$. For the parameters used, this gives $\epsilon_k \approx 13$ keV. This approximate expression has no dependence upon the parameter $\alpha$, so the limiting of $\alpha < B_s$ should not have a large effect on this result. 

\subsection{Fan Global Trajectories}

\begin{figure}
\centering
\subfloat[]{\label{fandist:a}
\includegraphics[width=0.5\textwidth]{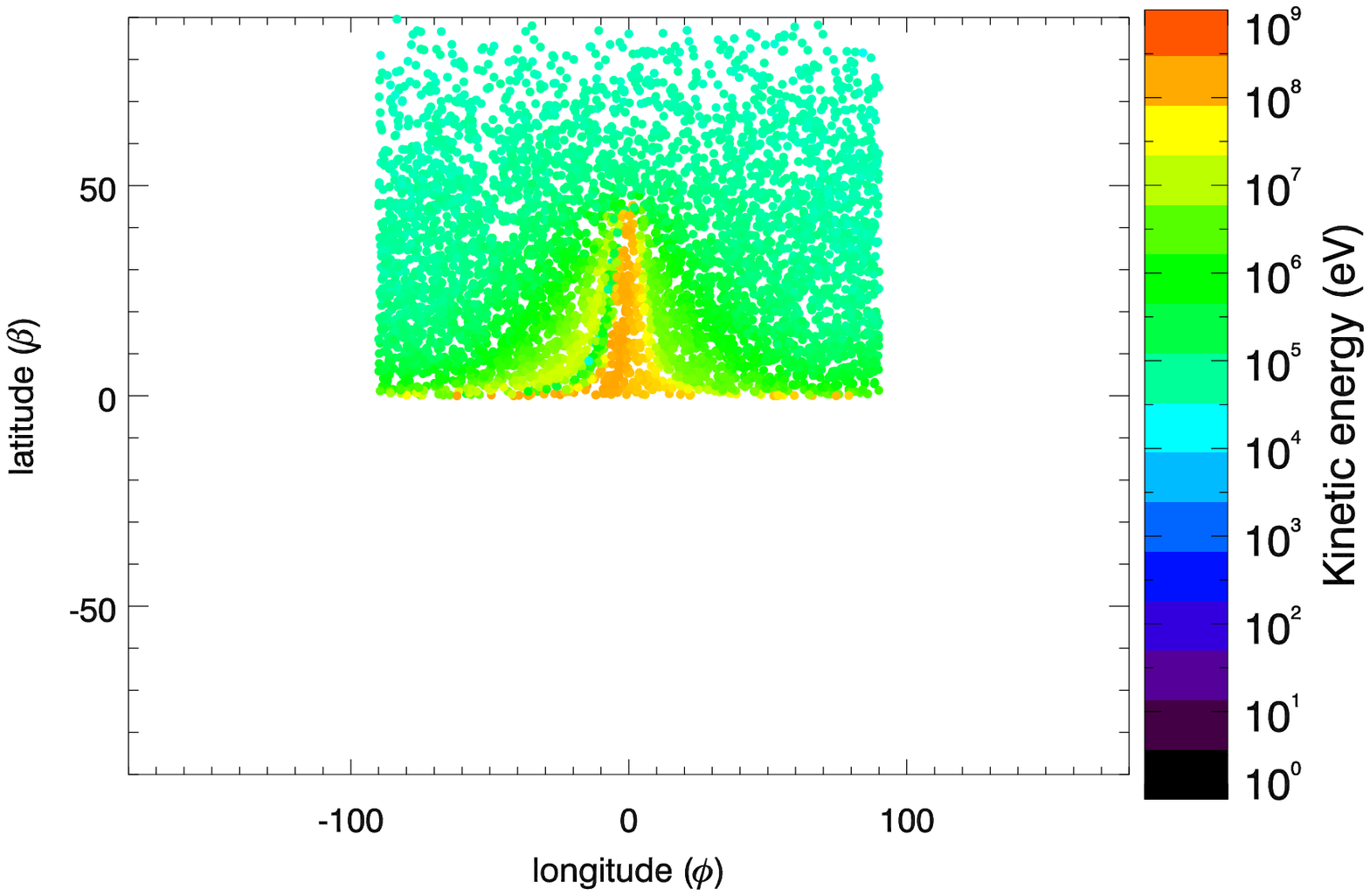}}\\[-0.2cm]
\subfloat[]{\label{fandist:b}
\includegraphics[width=0.5\textwidth]{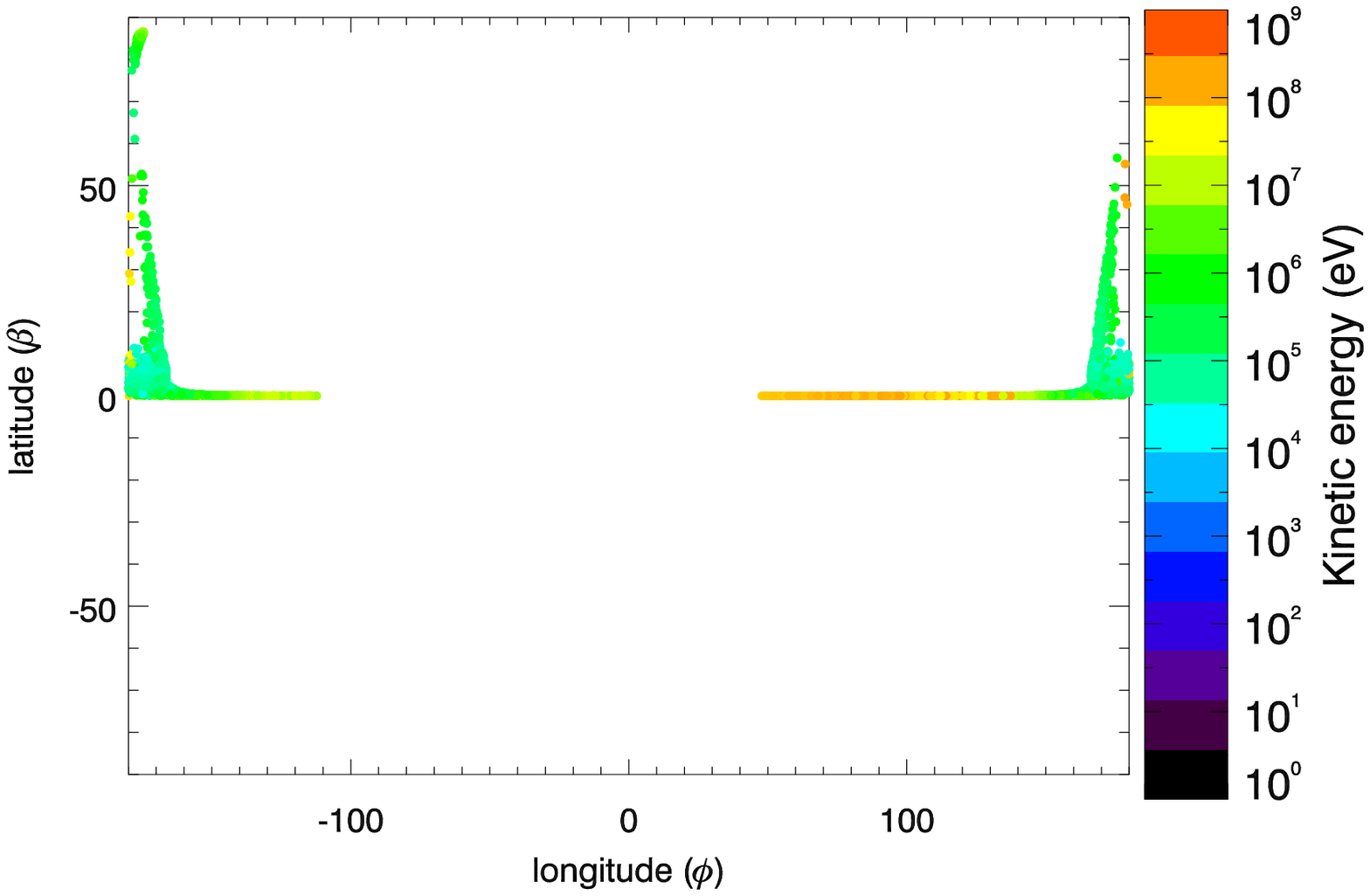}}
\caption{a) Angular distributions of protons in fan model at $t=0$, with initial temperature $T=86$ eV. The x-axis is $\phi=0$ and the fan plane is $\beta=0$. Protons are coloured by the final energy at $t=4\,000$. Parameters used are $\lambda = 0.75$, $B_s = \alpha = 10$, $\eta = 10^{-6}$. b) Angular distributions at time $t=4\, 000 T_{\omega,p}$ when the energy spectrum has reached steady state.}\label{fandist}
\end{figure}

The many particle simulation for the fan model is shown in Figure~\ref{fandist} for the optimised solution $B_s = \alpha = 10$, with $\eta=10^{-6}$, $\lambda = 0.75$. The initial distribution has thermal energy $\epsilon_k = 86$ eV with uniform random position in the upper inflow quadrant $-90^{\circ}<\phi<90^{\circ}$ and $0<\beta<90^{\circ}$. The final angular distribution is taken from when the proton distribution reaches a steady state in energy at $t=4000 \,T_{\omega,p} \approx 0.025$ s. This is more than two orders of magnitude faster than the spine model for similar parameters (even after the spine drift was increased by limiting $\alpha<B_s$) as the external electric drift~(equation~\ref{fanexternaldrift}) scales more favourably with the resistivity. Protons that cross the $R = 5L_0$ spherical boundary from the null point before this time are stopped and the energy and angular position at time of crossing is used. The $t=0$ angular distribution has some structure in terms of final energy gain as the initial random thermal velocities are dominated by the strong electric drift.

Within this structure there is some asymmetry in $\phi$. Indeed, we do not expect symmetry between particles drifting clockwise and anti-clockwise about the null as in the ideal case~\citep{db2} now that there is a current in the $y$ direction. Those protons with $\epsilon_k \approx 10^7$ eV at $\phi \approx -20^{\circ}$ (the yellow vertical band to the left of the green vertical band in Figure~\ref{fandist}\subref{fandist:a}) do not enter the current sheet, but gain high energy, as they are unmagnetised slightly, $\rho/L_{\nabla B} \sim 10^{-3}$, due to very fast electric drifts close to the sheet. Here the first adiabatic invariant, the constancy of $\mu$, is also violated.

Typically, the high energy protons of Figure~\ref{fandist} start either close to the x-axis at low to mid latitudes (about $7\%$ of the total number at latitude $\beta \gtrsim 1^{\circ}$ with final energy $\epsilon_{k,fin} \gtrsim 10$ MeV), or they start at very low latitude close to the fan plane ($<1\%$ of total at $\beta \lesssim 1^{\circ}$ and $\epsilon_{k,fin} \gtrsim 10$ MeV). At $t=4000\, T_{\omega,p}$ these energetic protons are found at $\beta \approx 0$ either side of $\phi=90^{\circ}$; the y-axis in the direction of the fan current.

 At $t=4\,000\, T_{\omega,p}$ there are a small number of high energy protons scattered at high latitudes (about $0.1\%$ with $\epsilon_k>10$ MeV). These enter the current sheet temporarily at negative longitude far from the null point, but exit again without any Speiser-like motion. They become slightly unmagnetised, with maximum $\rho/L_{\nabla B} \approx 10^{-2}$, following complicated trajectories. As they are not typical they are not investigated further in the external region, but the behaviour within the current sheet is discussed below (shown in Figure~\ref{fanCStrajgeneral}).

\begin{figure}
\centering
\subfloat[]{\includegraphics[width=0.5\textwidth]{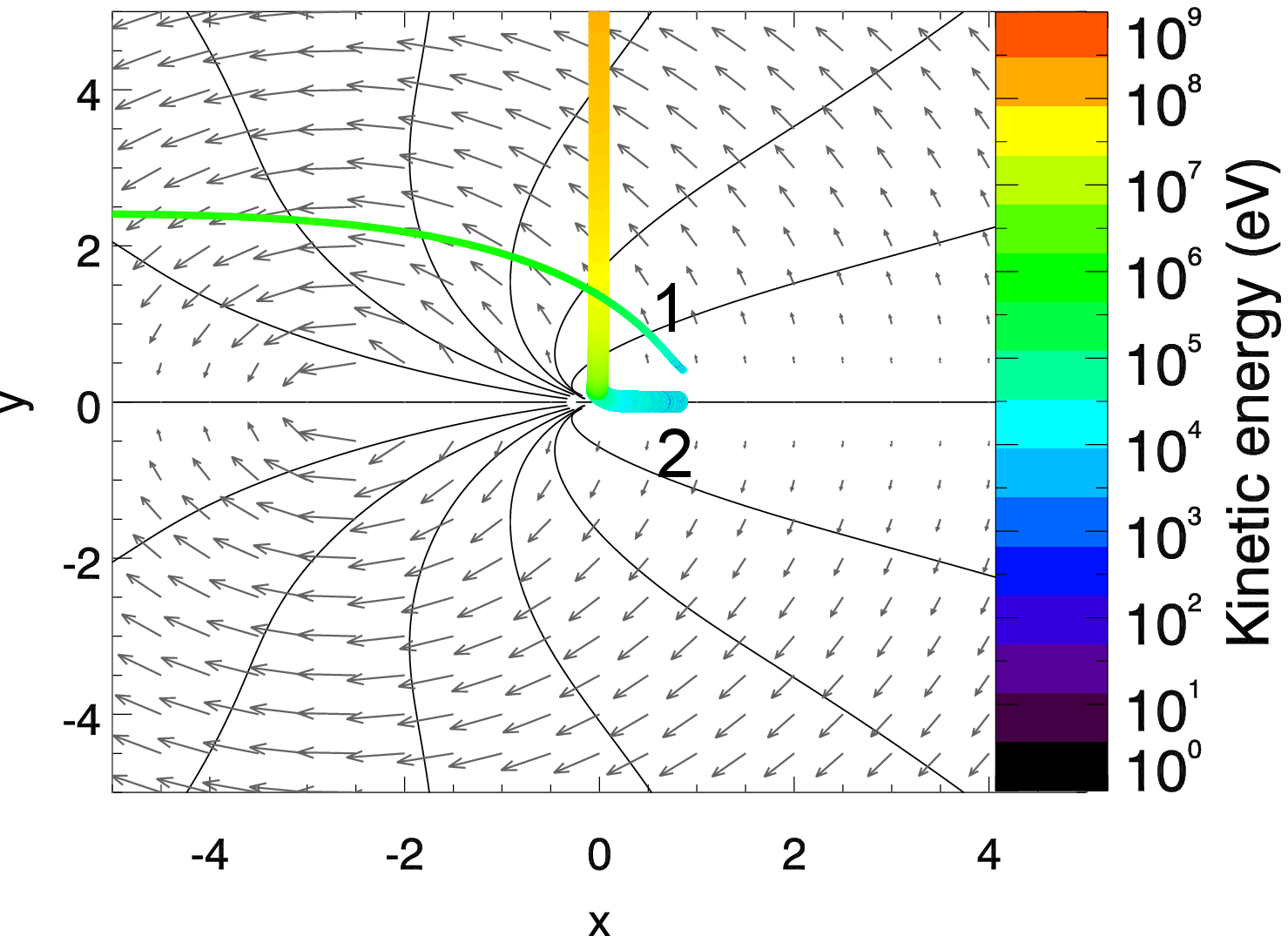}}\\
\subfloat[]{\includegraphics[width=0.5\textwidth]{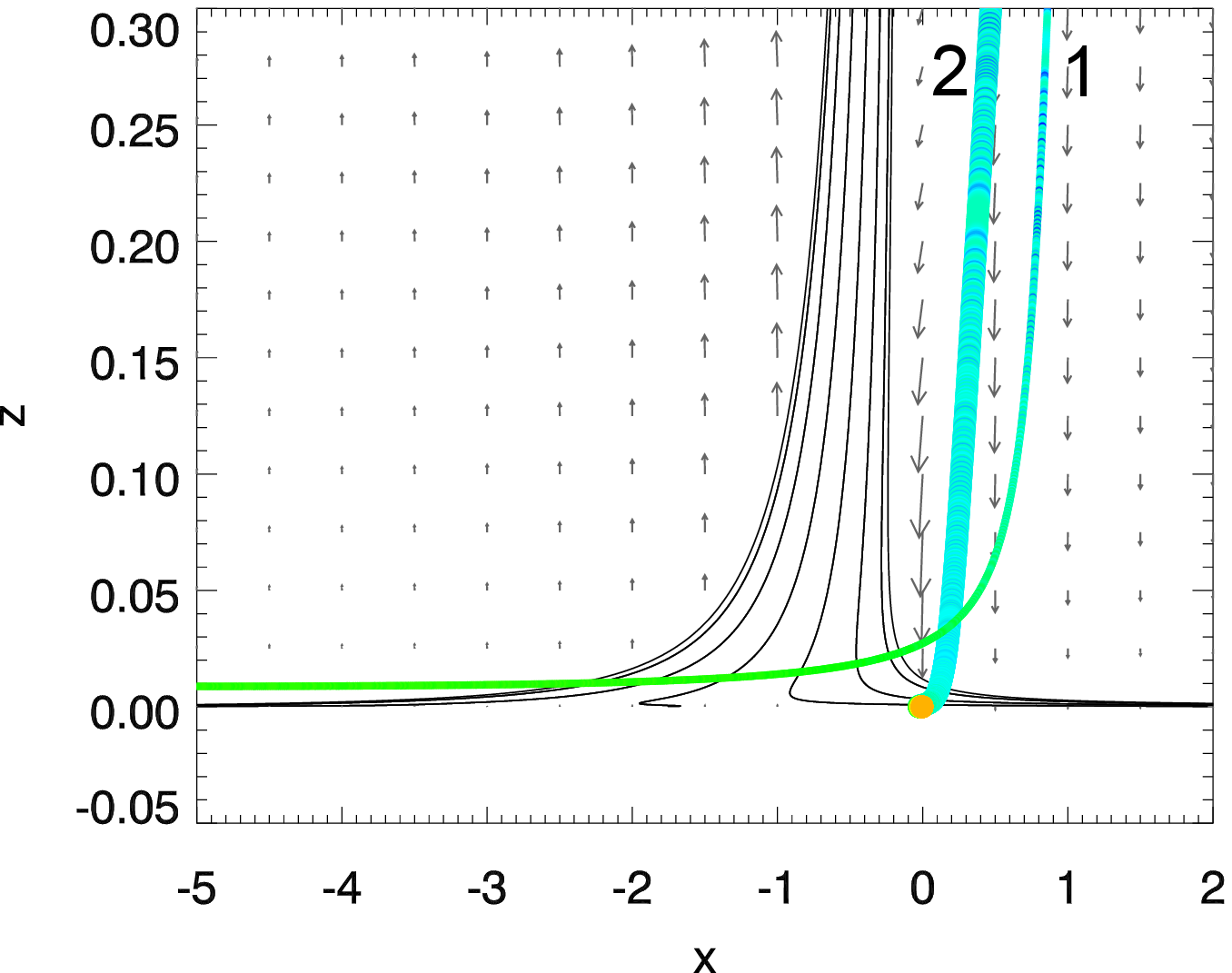}}
\caption{Two typical proton trajectories from the many particle simulations in the fan model. Proton '1' is represented by a thin line with initial position $(0.86,0.41,0.30)$, and proton '2' by a thick line with initial position $(0.8,0.003,0.6)$. The parameters are  $\lambda = 0.75$, $B_s = \alpha = 10$, $\eta = 10^{-6}$. The solid lines are representative magnetic field lines (seeded from the top of the spine axis and projected into the 2D planes)  and the arrows show the direction and relative magnitude of the electric drift velocity. a) In the x-y plane, where the electric drift arrows are from the edge of the current sheet $z=z_\eta$. b) In the x-z plane close to the current sheet, where the electric drift arrows are plotted on $y=0$. The initial positions are not shown in this plane.}
\label{fanTraj}
\end{figure}

Figure~\ref{fanTraj} shows the trajectory of two typical protons taken from the simulation. Proton '1' starts at $(x_0,y_0,z_0) = (0.86,0.41,0.30)$ and drifts around the null point due to the strong azimuthal electric drift. Although it drifts down towards the current sheet, it reaches a minimum height of $z \approx 15 \, z_\eta$ before it flows into the outflow quadrant, not entering the sheet. The main velocity contribution is electric drift as it moves around the null point, but $v_{\parallel}$ becomes dominant as the particle exits the simulation box parallel to the negative x-axis. The first adiabatic invariant is not violated, $\mu = const.$ and the maximum $\rho/L_{\nabla B} \sim 10^{-4}$ at closest point of approach to the sheet. The proton is strongly magnetised throughout. Despite not reaching the current sheet the energy gain is still considerable, reaching $0.5$ MeV as it crosses the $R=5$ sphere. 

Particle 2 starts at $(0.8,0.003,0.6)L_0$. The azimuthal electric drift is weak close to the x-axis and the proton drifts down to the fan current sheet. It enters the sheet at $(x,y) = (-0.02,0.15)$ and becomes unmagnetised: $\rho/L_{\nabla B}>1$ and $\mu$ is not conserved. We observe Speiser-like oscillations as the proton is accelerated in the y-direction. At $t = 0.846$ ms after entering the sheet, it passes out of the simulation box at $R=5$. Here, the particle is still within the sheet with $v_\parallel = 0.36 c$ and $\epsilon_k = 67$ MeV. Using this time period in the direct acceleration formula, $y = qE_0t^2/2m$, with the electric field on $z=0$, $E_0=\eta B_s/\bar{\eta}^{1/2} \quad [v_{Ae}B_0]$ from~(\ref{fanelec}), gives $y\approx 5$. Thus the proton is directly accelerated in the current sheet for the entire length of the simulation box. However, this motion is not Speiser-like throughout as $\rho/L_{\nabla B} < 10^{-2}$ when the proton reaches $y=1.5 L_0$. The proton reaches a global distance in the y-direction and becomes magnetised by the background $B_y$ component of the magnetic field, which acts as a kind of guide field. When the simulation is run without stopping the particle at $R=5$, the proton is not ejected from the current sheet throughout the whole simulation time $t=4\,000\,T_{\omega,p}$. 

This particle enters the current sheet at a distance $R\approx 0.15$ from the null point; however, this distance is not typical for the many particle simulation in Figure~\ref{fandist}. In the simulation 9.3\% of the total particles reach the current sheet, after a mean time of about $800 T_{\omega,p}$. The average distance from the null point of particles entering the sheet is $R \approx 2.2$; some remain magnetised by the background magnetic field inside the sheet.

\begin{figure}
\centering
\includegraphics[width=0.5\textwidth]{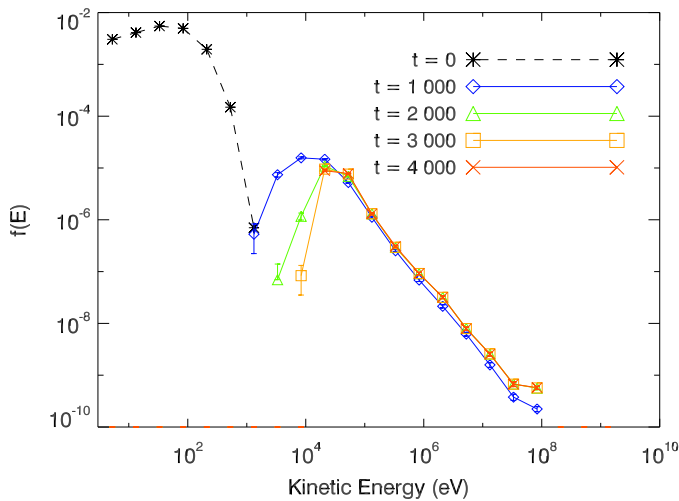}
\caption{Energy spectrum for the many particle fan simulation for protons with parameters $\lambda = 0.75$, $B_s = \alpha = 10$, $\eta = 10^{-6}$. For particles leaving the $R=5$ sphere, the energy at the time of crossing is used.}
\label{fanEnergyglobal}
\end{figure}

The energy spectrum for the fan simulation is shown in Figure~\ref{fanEnergyglobal}. Almost all the protons are accelerated into a non-thermal power law distribution $f(E) \propto E^{-\gamma}$, with slope $\gamma \approx 1.5$. For most particles, this efficient acceleration is due to the fast electric drift speed in the fan model being much larger than the initial thermal velocity. The spectrum appears to have reached a steady state by $t=4\,000 T_{\omega,p}$; however, it also depends upon the position at which protons are stopped as they leave the simulation box. As a test we repeat the simulation but stopping the protons at a spherical surface of radius $R=10$ from the null point, instead of $R=5$ that has been used consistently throughout these simulations. Now the 'flat tail' at $10^{7.5-8}$ eV in Figure~\ref{fanEnergyglobal} becomes a 'bump on tail' centred at $10^8$ eV (not shown) disconnected from the main distribution. Here, the power law part remains mostly unchanged. The population of protons that is trapped in the sheet as it crosses $R=5$ due to the strong 'guide field' remains trapped at $R=10$ where $B_y(y)$ has doubled in strength.

\subsection{Fan Current Sheet Trajectories}

The simulations considered thus far concern proton trajectories starting from the external region, at a distance $R=1$ from the null point. However, most of the protons entering the current sheet do so far from the null point. In the following, protons are initially distributed within the fan current sheet close to the null, to study the transition from non-adiabatic to adiabatic motion.

\begin{figure}
\centering
\subfloat[$t=2\,500$]{\label{fanCStraj:a}
\includegraphics[width=0.5\textwidth]{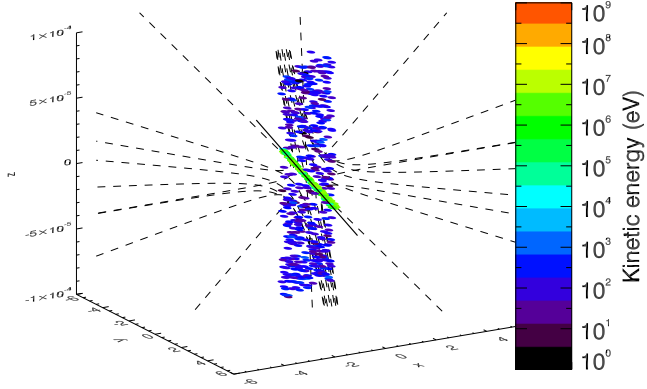}}\\[-0.2cm]
\subfloat[$t=17\,500$]{\label{fanCStraj:b}
\includegraphics[width=0.5\textwidth]{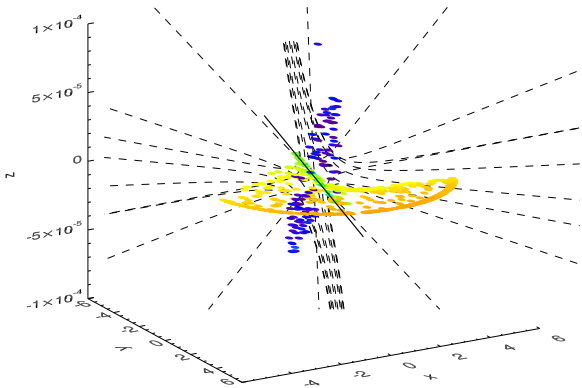}}
\caption{Proton positions after initial distribution within the fan current sheet, such that $|x_0|<1$, $|y_0|=0$, $|z_0| < z_{\eta}=\sqrt{2\bar{\eta}}$. Parameters used are $\lambda = 0.75$, $B_s = \alpha = 5$, $\eta = 10^{-8}$, $L_0 = 10^6$ m. The dashed lines are representative magnetic field-lines inside the current sheet (note the difference in scale of the z-axis). The solid black line is the line $(x_1,0,z_1)$ such that $B_x(x_1,0,z_1)=0$. The particles are stopped at $R=5$ from the null point.}
\label{fanCStraj}
\end{figure}

Firstly, we place particles within the sheet so that they are initially unmagnetised by the $B_y(y)$ component of the background field. They are magnetised only by the strong $B_x(x,z)$. The protons are uniformly distributed in the area $|x|<1$; $y=0$; $|z|<z_\eta$ with initial thermal energy $T=86$ eV. Figure~\ref{fanCStraj} shows the position of $2\,000$ protons at $t=2\,500\, T_{\omega,p}$ (a), and $t=17\,500\, T_{\omega,p}$ (b), during this simulation for the parameters $\lambda = 0.75$, $B_s = \alpha = 5$, $\eta = 10^{-8}$. We increase the dimensional box length to $L_0 = 10^6$ m as velocities in the current sheet are typically fast, giving reasonable integration times. This makes our results more comparable to the approximate analytic solutions of \citet{litvinenko06}. Note that $\eta$ decreases due to the increase in $L_0$ in equation~(\ref{eta}). We again artificially stop the particles as they cross the $R=5$ spherical surface.

At $t=2\,500\, T_{\omega p}$ most of the protons are strongly magnetised by the $B_x(x,z)$  magnetic field. Inside the current sheet, $|z|<z_\eta$ we can use equation~(\ref{powerseries}) to get approximate expressions for the electric and magnetic fields, 
\begin{equation}\label{taylorE}\boldsymbol{E} \approx E_y \boldsymbol{\hat{y}} \approx \eta \, B_s / z_\eta\boldsymbol{\hat{y}} \quad [v_{Ae}B_0],\end{equation}
\begin{equation}\label{taylorB}\boldsymbol{B} \approx \left(\frac{\lambda \alpha x}{2} + \frac{B_s z}{z_\eta}, \frac{\lambda \alpha y}{2}, -\lambda \alpha z\right)\quad[B_0],\end{equation}
$E_z$ is small except at global distance in $y$ (see below). 

For a proton starting at $x=0$, $y=0$, $z=z_\eta$, on the edge of the current sheet, the background components of the magnetic field are negligible. The proton drifts towards the vertical centre of the sheet $v_{Ez} \approx -(\eta/z) \, \boldsymbol{\hat{z}} \quad [v_{Ae}]$. It becomes unmagnetised at the fan plane, $z \approx 0$,  close to the null point and is directly accelerated in the y-direction. We compare this trajectory to the analytical WKB solutions of \citet{litvinenko06}. The ejection time for a non-relativistic proton that is unmagnetised close the null point, $x \approx 0,\, z\approx 0$, in the fan current sheet in our parameters is
\begin{equation}\label{tacc}t_{ejec} \approx \left(\frac{m^2 B_s L_0}{q^2\, z_\eta\, B_0 \lambda^2 \alpha^2 E_y}\right)^{1/3},\end{equation}
\citep{litvinenko06}~provided that the proton remains within the non-adiabatic region and the displacement magnetic field gradient is much stronger than the gradient from the background component, $B_s/z_\eta \gg \lambda \alpha$. The second assumption is valid for our simulation; however, we do not observe proton energy gain limited by ejection in these simulations. To understand this, we consider the distance travelled in the y-direction during this time, 
\begin{equation}y_{ejec} = y(t_{ejec}) \approx \frac{qE_y t_{ejec}^2}{2m},\end{equation}
which we compare with size of the non-adiabatic region from the null in this direction. The particle begins to be re-magnetised by the background field at a global distance $y^*$ such that
\begin{equation}v(y^*)/y^* \approx \omega_{B_y(y^*)}\end{equation}
 where $v(y)$ is a typical proton velocity and $\omega_{B_y(y)}$ is the gyro-frequency of a particle gyrating around $B_y(y)$. We use $v(y) = \left(2 q E_y y/m\right)^{1/2}$ from direct acceleration (if we use $v(y)=E_y/B_y$ the value for $y^*$ differs by $2^{1/3}$), assuming that there was no initial y-velocity and the particle entered the sheet at $y\approx0$. We recover the result of \citet{litvinenko06}, that in dimensional form
\begin{equation}\label{ystar}y^* \approx \left(\frac{8 m E_y}{q (B_0 \lambda \alpha)^2 L_0}\right)^{1/3}L_0.\end{equation}
The ratio of these two distances is
\begin{equation}y^*/y_{ejec} \approx \left(\frac{\lambda^2 \alpha^2}{B_s^2/z_{\eta}^2}\right)^{1/3},\end{equation}
where we have ignored factors of order unity. The ratio of the two timescales is the square root of this. There is little gyro-turning for protons starting close to the null point as this ratio is necessarily small for the fan current sheet. The proton is magnetised by the $B_y(y)$ ``guide field'' and trapped in the sheet, the energy gain is only bounded by the length of the sheet. 

Figure~\ref{fanCStraj} also shows the more general case of protons starting at $y=0$, $|z|<z_\eta$ and at a global distance in $x$. These protons drift vertically until they reach the diagonal line where $B_x(x,z)=0$, at which they become unmagnetised and accelerated. We do appear to see some gyro-turning for protons starting at $|x| \approx 1$. This is probably due to the strong component of the Lorentz force, $v_yB_z$, that acts to turn the trajectory to the x-direction. For particles starting at $x=0$ the $B_z$ magnetic field switches sign during the z-oscillations, but at $x=1$ the proton is unmagnetized below the fan plane and $B_z$ stays positive. The proton is turned in the positive x-direction but is quickly magnetised by the guide field when it reaches a distance of about $y^*$~(see equation~\ref{ystar}). In Figure~\ref{fanCStraj} it can be seen that particles are accelerated radially outwards from the null. They continue to gain energy as they become magnetised about the background field $\boldsymbol{P}$, on a field-line with a parallel component of the electric field $E_{\parallel}=\boldsymbol{E}\cdot \boldsymbol{P}/|\boldsymbol{P}|$.

\begin{figure}
\centering
\includegraphics[width=0.5\textwidth]{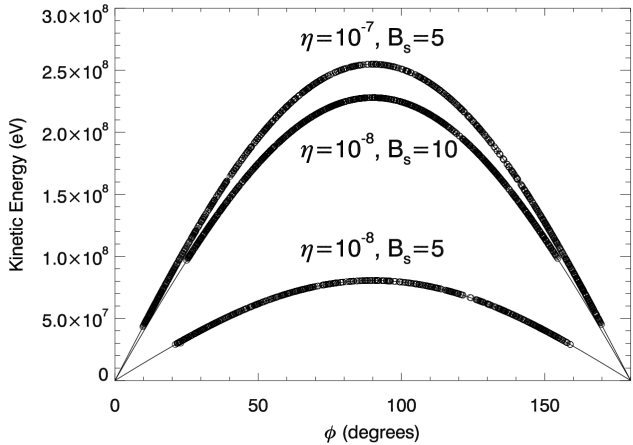}
\caption{Energy distribution of particles as they cross the $R=5$ boundary, where initial position is within $|x_0|<1$, $|y_0|=0$, $|z_0| < z_{\eta}=\sqrt{2\bar{\eta}}$. Here, $L_0=10^6$ m and the results for different values of $\eta$ and $B_s$ are plotted. The solid points are protons from the three simulations and the thin lines show the $\sin{\phi}$ relationship in equation~(\ref{sinenergy}).}\label{energyphifan}
\end{figure}

We artificially stop the protons at $R=5\,L_0$ from the null point. At $t=50\,000\,T_{\omega,p}$ all of the protons in the simulation have reached this distance without being ejected and we fit the energies of the particles by the expression 
\begin{align}\label{sinenergy}\epsilon_k (\phi) &\approx q E_{\parallel}(\phi)\, 5 L_0 \nonumber\\
&\approx 5\, q \, \eta^{1/2}\, B_s^{3/2} \, (1-\lambda^2)^{1/2}\sin{\phi}\quad [v_{Ae}B_0L_0],\end{align}
for the optimised solution $\alpha=B_s$, where $\phi$ is the azimuthal angle ($\phi=90^{\circ}$ is parallel to the current). Figure~\ref{energyphifan} shows the energies of $5\,000$ protons in three simulations with identical setup to Figure~\ref{fanCStraj} but with different values of $\eta$ and $B_s$. This expression (thin line) fits the energies of simulated particles (circles) as they cross $R=5\, L_0$ very well.  

\begin{figure}
\centering
\subfloat[$t=1000$]{\includegraphics[width=0.5\textwidth]{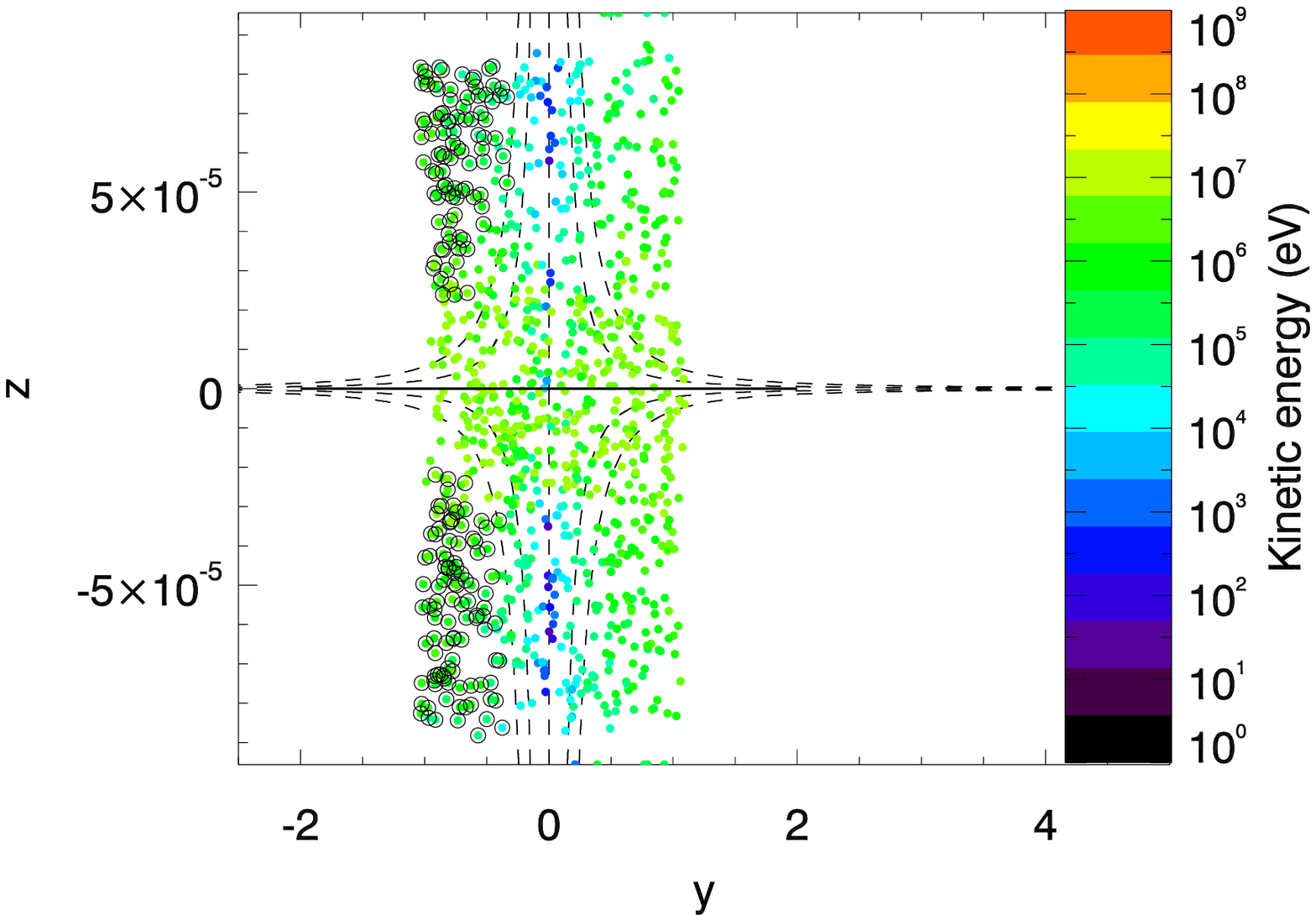}}\\[-0.2cm]
\subfloat[$t=8000$]{\includegraphics[width=0.5\textwidth]{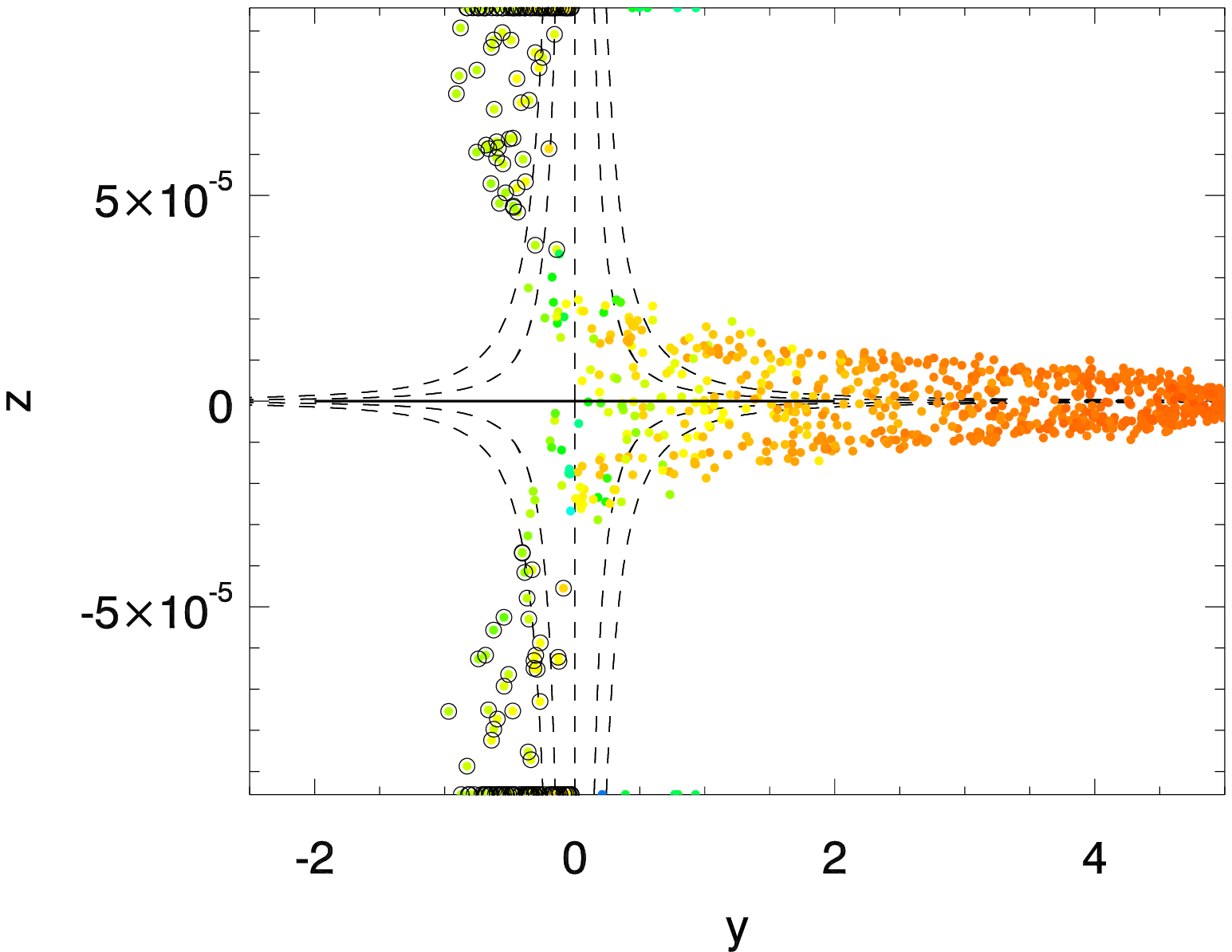}}
\caption{Proton positions after initial distribution within the fan current sheet such that $|x_0|,|y_0|<1$ and $|z_0| < z_{\eta}=\sqrt{2\bar{\eta}}$. Parameters used are $\lambda = 0.75$, $B_s = \alpha = 5$, $\eta = 10^{-8}$, $L_0 = 10^6$ m. Particles circled in black are those that start in $y<0$ and cross $z=z_\eta\approx 9.6\times 10^{-5}$ before $t=19\,000\,T_{\omega,p}$.}
\label{fanCStrajgeneral}
\end{figure}

In Figure~\ref{fanCStrajgeneral} we place $5\,000$ protons in the fan current sheet with initial position in $|z|<z_\eta$, $-1<x,y<1$ so that they are initially magnetised by the ``guide field'' $B_y(y)$. This is the more general case, as protons reaching the current sheet from the external region will not typically do so at $y\approx 0$. The protons that do not start close to $y=0$ are directly accelerated without the initial drift phase. By $t=19\,000 \, T_{\omega,p}$ all of the protons have left the simulation box; either through the $R=5$ boundary, or through the edge of the current sheet $|z|=z_\eta$. The particles that cross $|z| = z_\eta$ in $y>0$ start close to the edge and leave due to initial thermal velocity. However, those starting with $y<0$ are ejected from well within the current sheet. These protons ($19.7\%$ of total number) are circled in Figure~\ref{fanCStrajgeneral}. Typically they remain magnetised, with $\rho/L_{\nabla B}$ in the range $10^{-4} - 10^{-2}$. They are not ejected due to gyro-turning in the sense of~\citet{speiser65} as this requires non-adiabatic motion. The trajectories in the y-z plane seem to follow the magnetic field lines closely, although they have strong electric drift from the $E_z$ component of the electric field. Within the current sheet, $|z|<z_\eta$ the truncated power series in equation~(\ref{powerseries}) gives the z-component of the electric field from equation~(\ref{fanelec}) as
\begin{equation}E_z \approx \frac{B_s \alpha}{2\bar{\eta}^{1/2}}(1-\lambda^2)yz,\end{equation}
which is stronger than the current electric field~(\ref{taylorE}) for global $y$ and $z\ne 0$. However, it only contributes to strong electric drift (not shown) in negative x-direction for protons in the upper half of the sheet $0<z<z_\eta$, and in the positive x-direction for $-z_\eta<z<0$. This electric field also contributes to $v_{Ey}$ but this is dominated by the direct electric field acceleration.

The protons that are not ejected from the current sheet have an approximate sinusoidal dependence in kinetic energy gain, given by equation~(\ref{sinenergy}) (there is a thicker spread of points  about the predicted lines than in Figure~\ref{energyphifan} due to differences in initial potential energy). 

\subsection{Scalings}

In the fan global simulation of Figure~\ref{fandist} we chose the optimised parameters $\eta=10^{-6}$, $B_s=\alpha=10$ and $\lambda=0.75$. With consideration to the large variation in both scale and behaviour in a given distribution of flares, it is interesting to see how the results of this simulation scale when the simulation parameters are varied.

\begin{figure}
\centering
\subfloat[]{\includegraphics[width=0.5\textwidth]{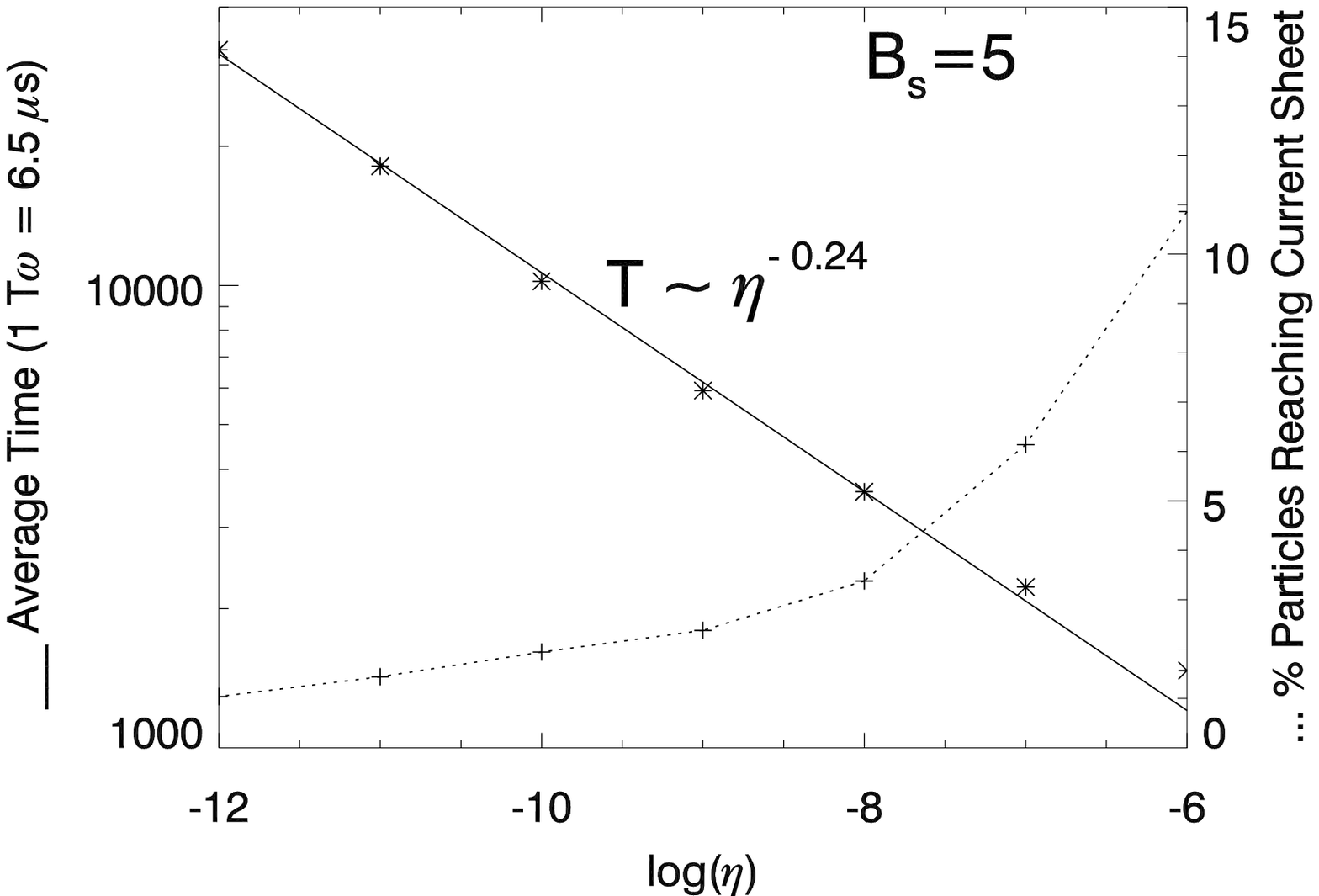}}\\[-0.22cm]
\subfloat[]{\includegraphics[width=0.5\textwidth]{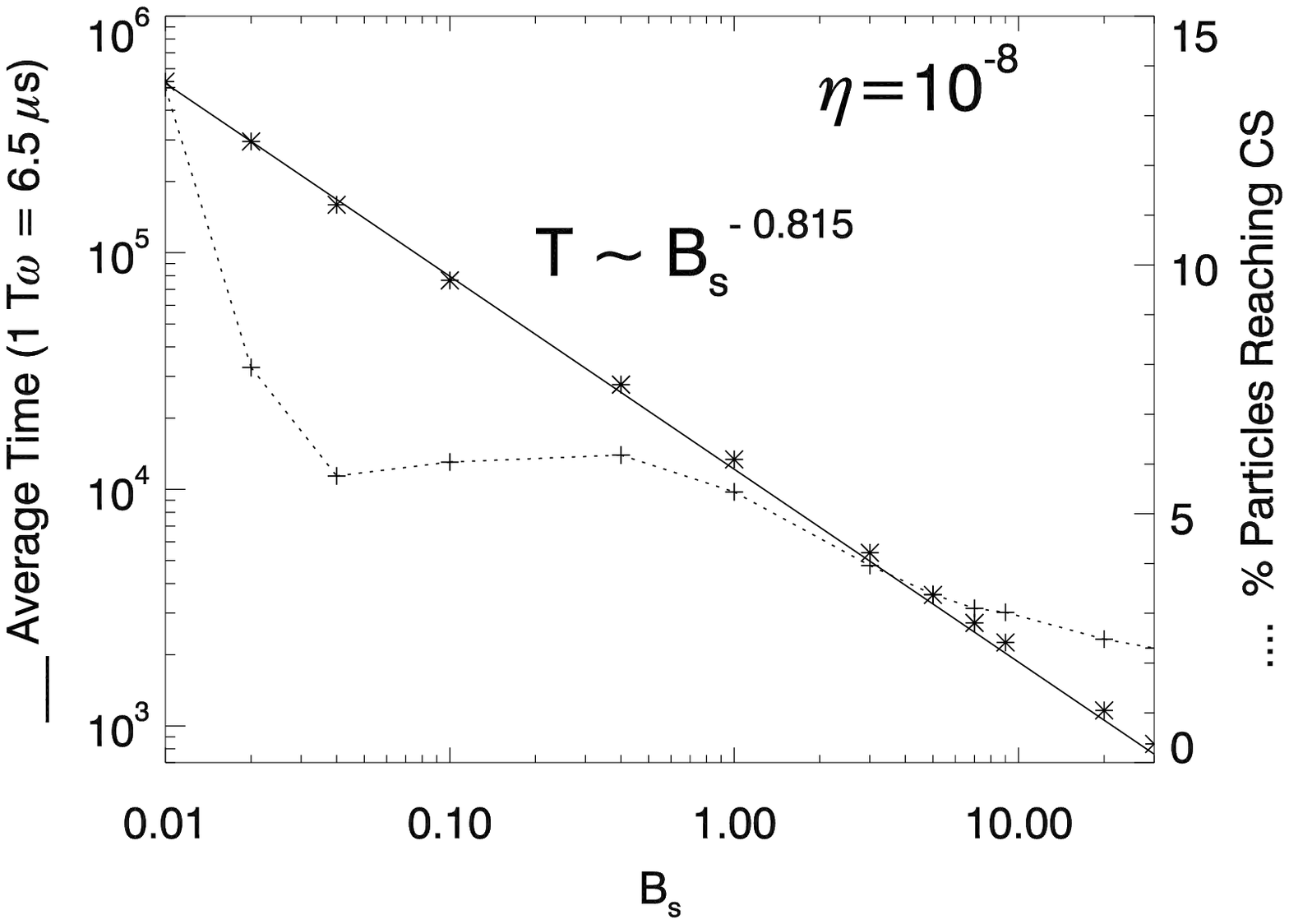}}\\[-0.22cm]
\subfloat[]{\label{fanTscaling:c}
\includegraphics[width=0.5\textwidth]{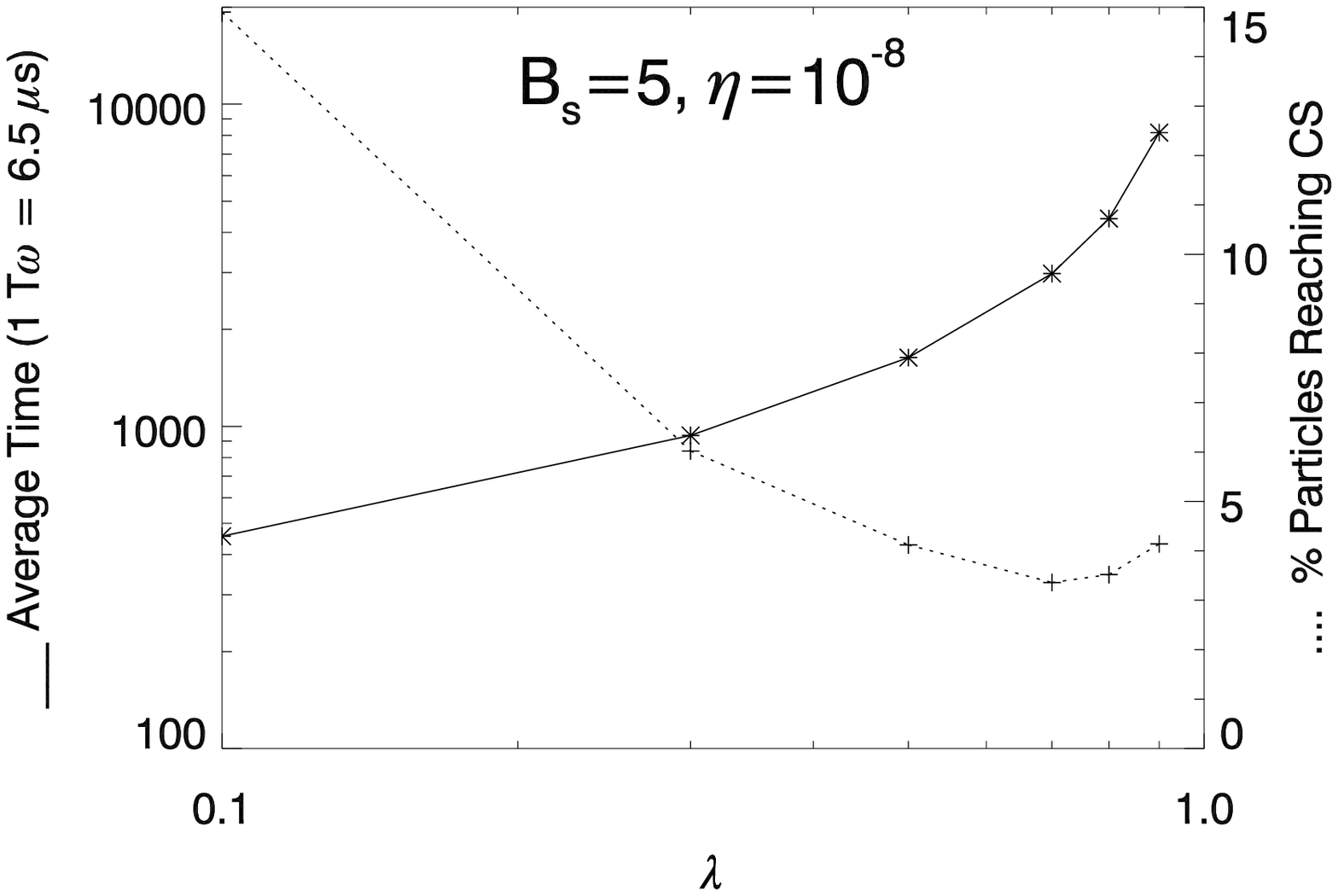}}
\caption{Percentage of total particles (+) reaching current sheet at $z=z_\eta$ from the external region ($R=1$ in the inflow quadrant), and the mean time taken (*), for different values of $\eta$, $B_s$ and $\lambda$. Each data point is from a many particle simulation with initial Maxwellian distribution ($T=86$ eV) of $5\,000$ protons. The set-up is the same as that in Figure~\ref{fandist}. a) For different values of $\eta$ with fixed $B_s=\alpha=5$, $\lambda=0.75$. The solid line is a least squares fit to the points. b) For different values of $B_s$ (with $B_s=\alpha$) for fixed $\eta=10^{-8}$, $\lambda=0.75$. The solid line is a least squares fit. c) For different $\lambda$ with fixed $B_s=\alpha=5$, $\eta=10^{-8}$. No curve was fit.}
\label{fanTscaling}
\end{figure}

The current sheet within the fan model is the most effective way to accelerate the particles. Thus, it is interesting to study the effect of varying parameters on the fraction of protons that enter the current sheet from the external region, and the average time taken to drift there from an initial position on the $R=1$ sphere. These scalings are shown in Figure~\ref{fanTscaling}. They are from simulations of $5\,000$ protons at $T=1$ MK starting at the upper inflow region at $R=1$. We define the current sheet as $z = z_\eta = \sqrt{2\bar{\eta}}$ for the fan model, although we note that not all of the protons reaching this height become non-adiabatic. 

\begin{figure}
\centering
\subfloat[]{\includegraphics[width=0.48\textwidth]{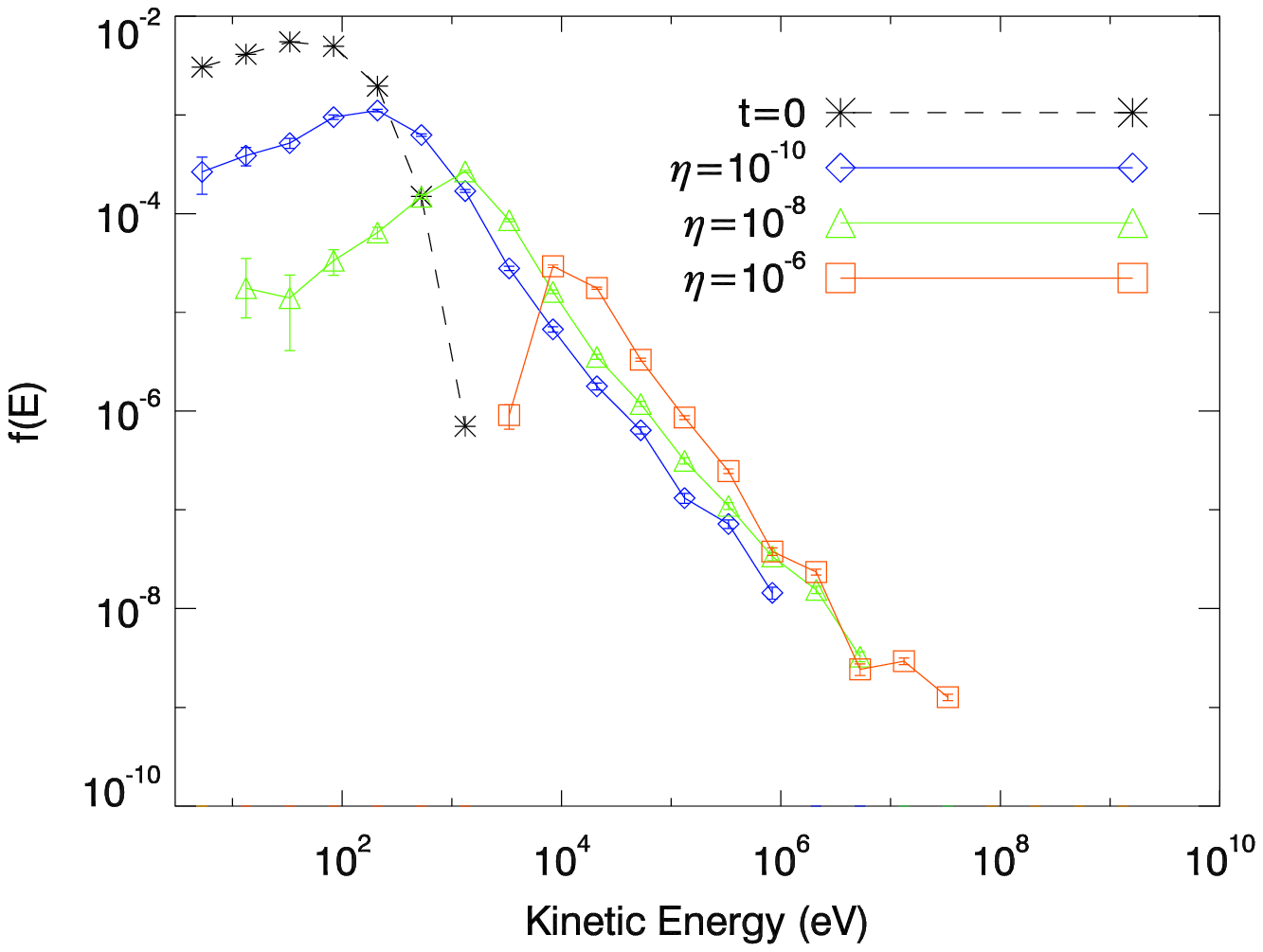}}\\[-0.22cm]
\subfloat[]{\includegraphics[width=0.48\textwidth]{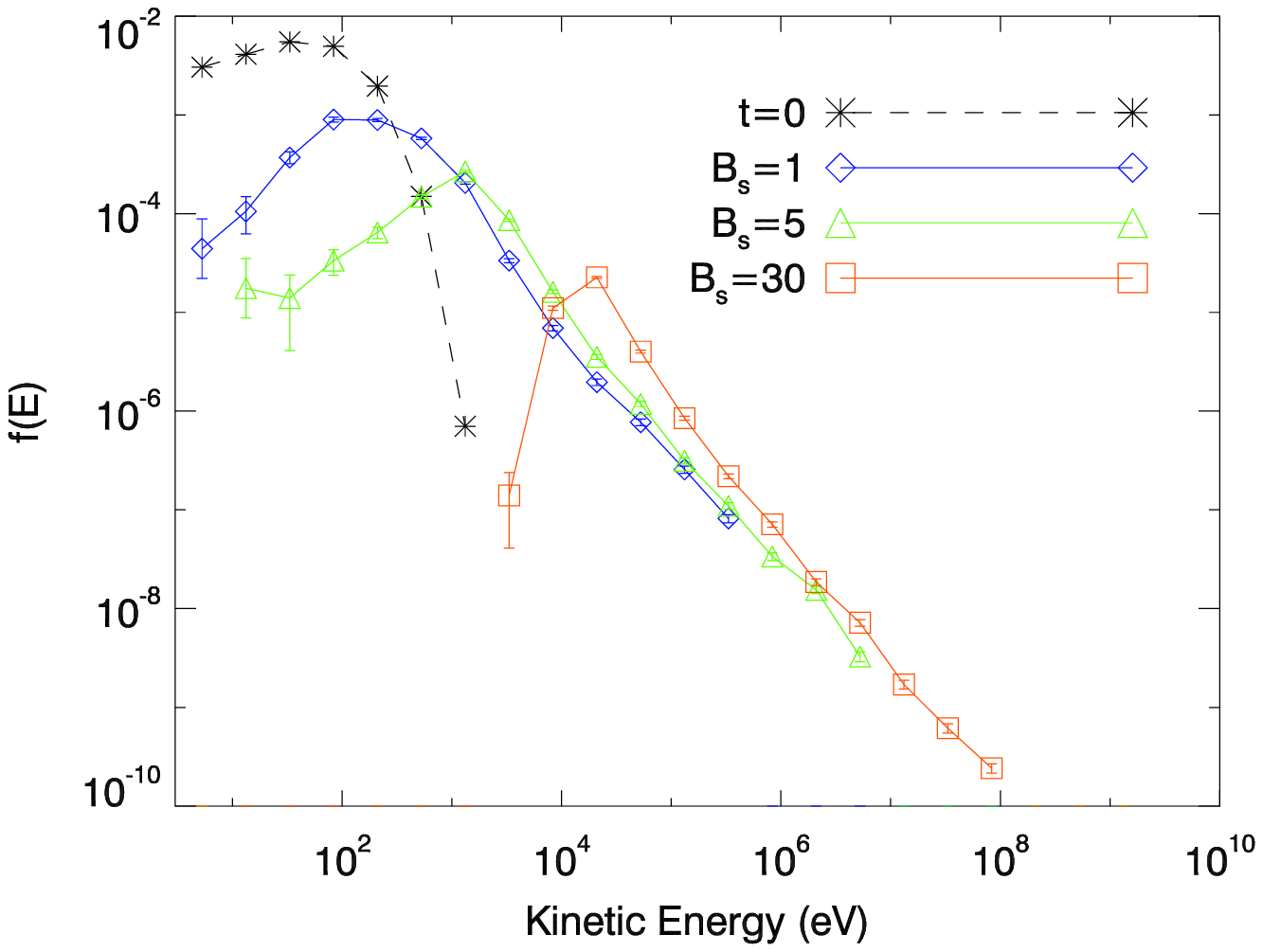}}\\[-0.22cm]
\subfloat[]{\label{enscaling:c}
\includegraphics[width=0.48\textwidth]{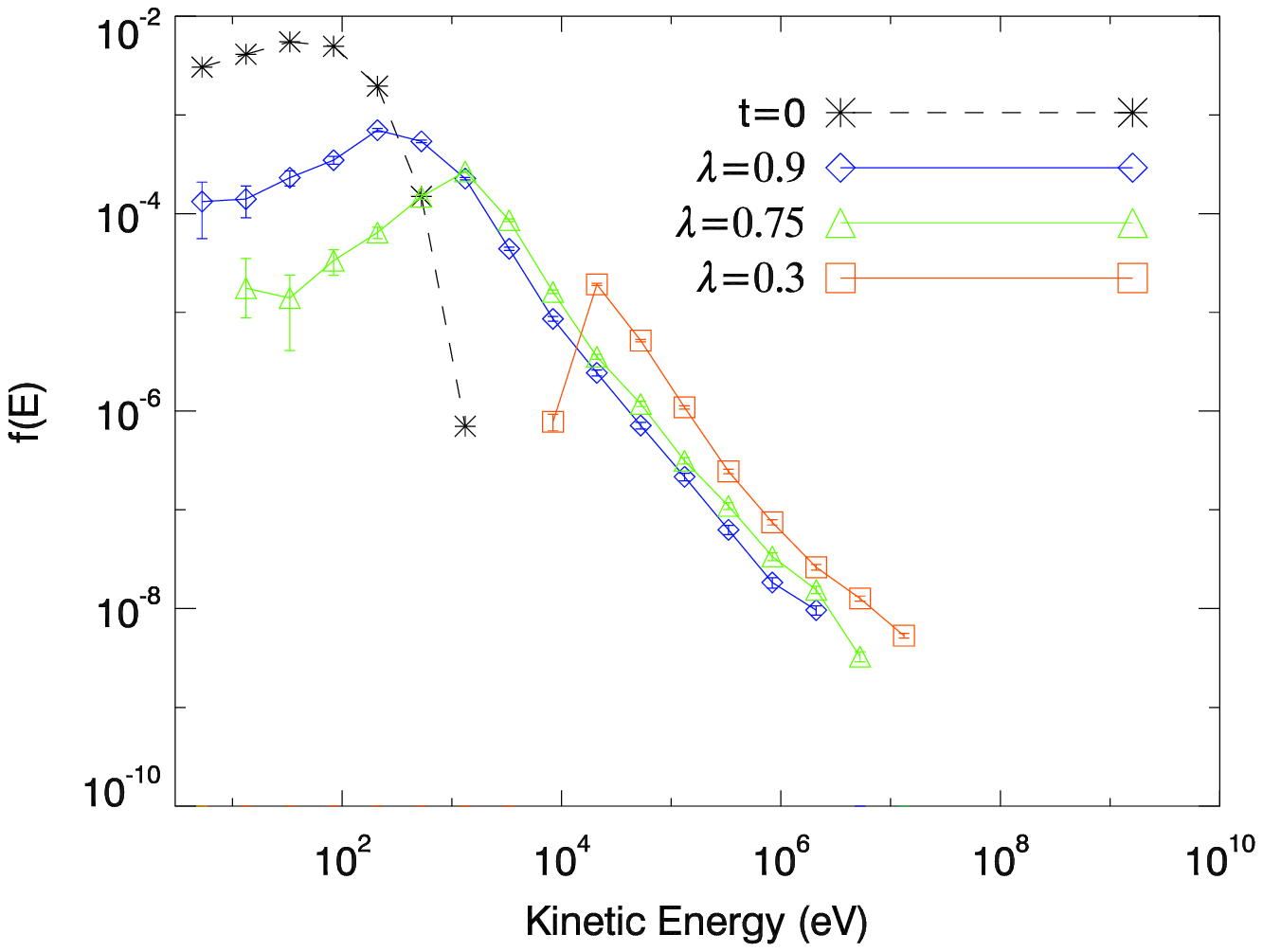}}
\caption{Scalings of steady-state energy spectra for global fan simulation. a) For different values of $\eta$ with fixed $B_s=\alpha=5$ and $\lambda=0.75$. The time taken to reach steady state was $t=8\times 10^4$, $t=2 \times 10^4$ and $t=6.4 \times 10^3\, T_{\omega,p}$ for $\eta=10^{-10}$, $\eta=10^{-8}$ and $\eta=10^{-6}$ respectively. b) For different values of $B_s=\alpha$ with fixed $\eta=10^{-8}$ and $\lambda=0.75$. Steady-state was reached at $t=2\times 10^5$, $t=2\times 10^4$ and $t=5 \times 10^3 \, T_{\omega,p}$ for $B_s=1$, $B_s=5$ and $B_s=30$ respectively. c) For different values of $\lambda$ with fixed $B_s=\alpha=5$ and $\eta=10^{-8}$. Time to steady state is $t=10^5$, $t=2 \times 10^4$ and $t=8 \times 10^3 \, T_{\omega,p}$ for $\lambda=0.9$, $\lambda=0.75$ and $\lambda=0.3$ respectively. }
\label{enscaling}
\end{figure}

The average time taken for the particles to reach the current sheet gives a measure of the external electric drift speed. The approximate drift scaling of equation~(\ref{fanexternaldrift}), $v_E \sim B_s^{3/4} \eta^{1/4}$, is in reasonable agreement with these drift times.

The fraction of particles reaching the sheet increases typically with increasing $\eta$ and decreasing $B_s$. Note that the size of the current sheet which we use to produce the scalings has the dependence $z_\eta \sim \eta^{1/2} B_s^{-1/2} (1-\lambda^2)^{-1/2}$, although this does not fully explain the result as the trends are not simple power laws. Figure~\ref{enscaling} shows how the energy spectra vary with these parameters. As might be expected the spectra shift to the right for an increase in both $B_s$ and $\eta$. Both the convective electric field (and so external electric drift) and the direct electric field within the sheet increase with larger values of these parameters.

Up until now we have used $\lambda=0.75$ as a constant in all of the simulations. This parameter has been typically left as constant in the calculation of MHD energy dissipation scalings~\citep{CFW97,cw2000} for the fan model as it can only be varied within an order one range. However, it has a large effect on the efficiency of the fan model for particle acceleration (see Figure~\ref{fanTscaling}\subref{fanTscaling:c} and Figure~\ref{enscaling}\subref{enscaling:c}). Varying $\lambda$ within $0\le \lambda<1$ has a comparable effect on the fraction of particles reaching the current sheet as varying $\eta$ by six orders of magnitude. Also, as shown in Figure~\ref{enscaling}, decreasing $\lambda$ shifts the energy spectrum to higher energy and decreases the time taken to reach steady-state. These effects can be explained somewhat by an increase in external drift speed, although varying $\lambda$ also has an effect on other quantities such as the current sheet height.

\section{Summary and Discussion}\label{secdiscussion}

We investigated test particle motion in the electromagnetic fields of~\citet{craig95,CF96} and \citet{CFW97}, that are solutions to the steady-state, incompressible and resistive MHD equations at a 3D null point. The study was carried out by modifying the code of~\citet{db1,db2,db3} and \citet{db4}. We considered initially Maxwellian ($T=86$ eV) distributions of protons starting at a global distance $R=L_0$ from the null point in resistive spine reconnection, where the electric current is within a thin cylinder about the spine axis, and resistive fan reconnection, with a current sheet in the fan plane. When the energy spectrum from the simulations reached steady-state we find the final angular position of the particles from the null and their energy distribution. We identified different populations of accelerated particles and, to understand the acceleration mechanism, examined a typical single particle trajectory in each case. For the fan model we ran additional simulations with the particles initially distributed within the fan current sheet, to study the effect of the null point on directly accelerated particles. We consider two cases, where particles are firstly unmagnetised and secondly magnetised in their initial position by a background ``guide field''. Finally, we show how the external drift times and energy spectra for the fan model scale when treating $\lambda$, $B_s$ and $\eta$ as free parameters (for the optimised solution $\alpha=B_s$, corresponding to the thinnest current sheet~\citep{cw2000}).

We found that the spine model, which gave promising acceleration results in the ideal case~\citep{db3}, is much less effective when resistive effects are included (at least for this specific resistive model). The electric drift in the external region is weak, scaling with resistivity as $v_E(r \gg r_\eta) \sim \eta$, giving very long drift times for protons to reach the spine axis. We find that there are two populations of accelerated particles. One of these escapes the simulation box down the base of the spine axis, similar to the proton jet found in the ideal model~\citep{db2}, and the other is close to the fan plane, where particles have crossed the current sheet in the spine axis. The energy gain for particles that reach the current sheet is low. The main limiting factor is the small electric potential energy difference across the current sheet, due to localisation of the reconnection electric field to a small cylinder about the spine axis. The apparent contrast with the ideal results arises from parameter choice. In the ideal regime, the magnitude of the electric field for the spine and fan models was set to equal strength in the external region, at a global distance from the null. The electric field falls steeply as $1/r$ in the external region of the spine model which gives very strong acceleration close to the spine in the ideal case. However, when the pressure constraints in the resistive model are taken into account, namely the limiting of the displacement field on the edge of the current sheet to avoid unphysical magnetic pressures~\citep{CFW97}, this $1/r$ dependence gives very weak electric field, and slow drift, in the external region.

We found much higher proton energies in the resistive fan model for similar parameters. For $\eta=10^{-6}$, $B_s=\alpha=10$ and $\lambda=0.75$ we find the energy spectrum from a distribution of protons starting with thermal energy at $R=L_0$ from the null becomes power law at steady state, with a spectral index of about $-1.5$ and maximum particle energy of the order $0.1$ GeV. The electric drift in the external region is much quicker than the spine model, $v_E \sim \eta^{1/4}$. It accelerates all of the particles in the simulation as it is faster than the initial random velocities associated with thermal motion at $T=86$ eV. We find that the population with the highest energy gain corresponds to protons that have entered the fan current sheet. The energy gain for these protons is not limited by ejection due to unstable motion as they are re-magnetised within the current sheet by a ``guide field''. The upper bound in energy gain is only limited by the electric potential energy, determined by the length of the current sheet. However, we find that a number of protons that enter the current sheet upstream of the null point can be ejected while remaining magnetised. This is due to the geometry of the background field lines, namely that they diverge at the null. We will study this effect in the future when we consider electrons. \citet{db4}, and \citet{guo} show that electrons remain magnetised at a closer distance to the null point, which may give a difference between the number of electrons and protons ejected in this manner.

We find that the parameter $\lambda$, which gives the degree of shear between the magnetic and velocity fields (such that $0 \le \lambda <1$) has a large effect on the final energy spectrum of protons in the fan model. In the limit of $\lambda = 0$ the magnetic field in the fan model is annihilated. As we expect magnetic field to still exist in the reconnection site after a topological change it would be more likely that $\lambda \approx 1$.

In these simulations we have neglected the electromagnetic effects of the non-thermal particles onto the background fields. This is typical of the test particle approach, where it is assumed that the number of particles in the current sheet is a small fraction of the total number. For a large range of parameters in the fan model (see Figure~\ref{fanTscaling}) this fraction is typically less than $5\%$ of the total number of particles starting in the inflow region. To estimate the strength of the magnetic fields from these particles, and the polarisation electric field from any charge seperation, it is necessary to also consider electrons which we will do in future work. A fully self-consistent approach, eg. using Particle In Cell simulations, is computationally expensive at present, particularly in fully 3D geometries due to the large dynamic range of spatial scales. We have also neglected compressibility, a simplification used to get the analytic solutions~\citep{CFW97}. It would be interesting in future to include both time-dependence and resistivity, by using electromagnetic fields from MHD simulations, to see how particles behave in so called~\textit{spine-fan} reconnection~\citep{pontin07a,pontin07compress}.

\bibliographystyle{aa}
\bibliography{paper}

\end{document}